# Remote Surgery with 5G or 6G: Knowledge Production and Diffusion Globally and in the German Case

Marina Martinelli[1]
André Tosi Furtado[1]
[1] – Universidade Estadual de Campinas (UNICAMP), Institute of Geosciences, Department of Science and Technology Policy

## ABSTRACT

This paper is a comprehensive exploring of technology capability in 5G/6G TIS, explicitly focusing on the potential of remote surgery globally and in Germany. The paper's main contribution is its ability to anticipate new debates on the interplay between TIS and contexts, with particular emphasis on the national and international levels. Our findings, derived from a Bibliometrics study of industry-academic relationships, highlight crucial collaborations in Germany, positioning the country as a strategic actor in international TIS and, by extension, in applying 5G/6G technological systems to remote surgery due to its knowledge production capability. We propose policies that can stimulate interaction between smaller suppliers and larger companies, which can act as intermediaries and provide access to international markets.

# 1. Introduction

5G and 6G networks are essential technologies to critical disputes, such as those in the semiconductor industry, transistors and integrated systems that lead to software and hardware construction - main components of cutting-edge networks (5G and 6G) (Rikap & Lundvall, 2021; Moldicz, 2021; Liu et al., 2018; Liefner et al., 2019). The Metaverse, with its immense potential, is set to revolutionise the healthcare technological system. The transformation will reshape medical education, improve patient care, and facilitate innovative medical research and treatment advances. As it continues to evolve, the integration into healthcare will undoubtedly redefine the future of medicine, offering a more personalised, accessible, and effective healthcare system for patients and professionals. This evolution is expected to bring substantial revenue benefits of USD$58 billion in 2030 and USD$99 billion in 2032 (Dhanalakshmi et al., 2024).

This article will demonstrate, through the Production and Diffusion of Knowledge, China's capacity to generate patents quantitatively. It will also highlight the significant contributions of other companies and countries, despite their smaller numbers, to the generation of patents crucial to the market, with a more significant citation impact. The publications aim to map the actors of the Technological Innovation System in 5G/6G, revealing nuances about the specific context of Germany.

Freeman & Perez (1988) understand the TIS in a meso perspective - at the national level - and in techno-economic terms, that affect directly and indirectly a conjunction of enterprises and industries. In terms of perspectives, micro refers to the local level, meso to the national, and macro to the international or global level. Furthermore, the “function” perspective - instead of motors, drivers, or mechanisms - describes the main generating and disseminating technology process (Furtado et al., 2020). Nevertheless, according to Carlsson & Stankiewicz (1991), “the boundaries of technological systems do not coincide with national boundaries”. Therefore, understanding of *National TIS* and *National Innovation System* differs because of the technological element, which comprises network formation, technology transfer, and spillover effects.

Scientific publications about developing nations underscore the successful assimilation of international technological transfers occurs only when accompanied by concurrent indigenous technological endeavours. Consequently, our focus accentuates the significance of domestic stakeholders and networks in advancing novel technologies on the strategic initiatives of local enterprises or organisations to integrate technology sourced from multinational corporations (Furtado et al., 2020; Liu et al., 2018).

Liu and colleagues (2018) attempted to conduct robust research on the 3G to 5G transition qualitatively from a socio-technical perspective for TIS. Their objective was to explore the evolution of this transition (3G — 5G) according to the telecommunication industrial efforts in developing indigenous technology for technology transfer. Their methodology contribution is particularly innovative in the 5G and TIS fields, as it has been divided into three steps: the three TIS 3G, 4G, and 5G.

Gonzalo et al. (2023) contribute to the topic of 5G from the perspective of the countries of the BRICS economic bloc and with a qualitative approach under the lens of National Innovation Systems. These authors mention 5G as a disruptive technology capable of placing India as a protagonist country in geopolitical disputes. The more geopolitical and penetrating approach places India, Brazil, and developing countries on a kick towards the binary geopolitical dispute. 5G and 6G are two complementary Technology Innovation Systems (TIS), one an evolution of the other, providing fast and ultra-fast Internet with low latency and ultra-low latency (Noor-A-Rahim et al., 2022). They represent different levels of high-speed Internet but are all part of the same entity. The complexity of actors, institutions, networks, and technologies involved in 5G enhances its status as a Technological Innovation System. It is a complex and heterogeneous system fundamental to driving technological change and on which such change depends.

Therefore, ***this research aims*** to analyse the Technological Innovation System (TIS) conceptual background applied to 5G/6G networks for remote surgery empirical case. This represents, in fact, an interconnection or interplay of two TIS: (i) 5G and 6G, and (ii) remote surgery. Applied to remote surgery medical empirical cases, 5G and 6G may be composed of a systemic interconnection of innovation systems. Even though this will be discussed in Section 3, the research question is:

***RQ:*** *How is the 5G/6G TIS capability concerning the applicability of remote surgery globally and in Germany?* This is of utmost importance in technological innovation and strategic healthcare because of (i) the tentative resolution of geographical context problems and (ii) the interplay of TIS from a systemic-structural viewpoint.

This article is organised in two phases: (i) a conceptual and methodological discussion and (ii) an analysis of the empirical case through software and Bibliometrics, exposing the essential data. The gaps identified lead to a theoretical discussion based on the TIS conceptual elements, addressing the disclosed in a 5G to 6G route correction. As an evolution, 6G is expected to enhance the 5G mobile communication system through innovations that will benefit it. Section 2 discusses the conceptual background elements. In section 3, we delve into methodological aspects on Bibliometrics analysis, considering patents and publications through software. Section 4 presents samples with results. In section 5 we discuss the literature and some conceptual background aspects. Finally, we conclude with our findings, future studies direction, and limitations.

## 1.2. 5G, 6G and Remote Surgery

The advancement of 5G is of crucial importance for strategic actors in TIS, such as Germany, especially for underprivileged populations and remote areas without adequate access to health services. 5G offers high speed and low latency, enabling real-time remote medical consultations, even in areas with limited infrastructure. This reduces the need for travel and improves preventive care, such as telemedicine. Regarding remote surgeries, 5G's low latency can enable remote surgeries, with specialist doctors operating on patients in distant areas. This innovation is only possible with 5G networks due to the delay in data transmission. Another critical point is health monitoring; 5G also facilitates using Internet of Things (IoT) devices, such as sensors, to monitor health conditions in real-time, allowing rapid interventions in critical cases. Furthermore, improving local care is possible with more excellent connectivity, providing online training for health professionals, and raising awareness among the population about primary healthcare (Elendu et al., 2024; Georgiou et al., 2021).

5G enables a wide range of services that were impossible with previous generations of networks. This is due to low latency, high speed and greater simultaneous connection capacity. Some primary services that 5G enables include Augmented Reality (AR), Virtual Reality (VR), and Extended Reality (XR) which can be used in real-time with no noticeable delay. In addition to deepening the use of AR, it becomes essential to use XR for remote surgery (Morimoto et al., 2022). This is valuable for areas such as entertainment, education and industrial training. On the other hand, Autonomous Vehicles enable real-time communication between vehicles and infrastructure, which is essential for the safety and operation of self-driving cars and ambulances, for example. The Internet of Things (IoT) on a large scale allows connected devices, such as industrial, agricultural or healthcare sensors, to communicate efficiently, promoting the development of intelligent and automated solutions. Imaging exams, such as “MRIs and CT scans”, can be transmitted in real time to specialists anywhere in the world. This enables fast and accurate diagnoses, improving the efficiency of medical treatments. These services are already available in some regions where 5G infrastructure has been deployed, particularly in advanced hospitals (Ramraj Dangi et al., 2021; Georgiou et al., 2021; Tychola et al., 2023).

Millimetre waves, operating at higher frequencies (above 24 GHz), offer high bandwidth, enabling high-speed data transmissions. The technology's advantages, including high data rates, are made possible by the millimetre wave band's high transmission capacity, which can deliver gigabits per second (Gbps) speeds, ideal for services that require high bandwidth, such as 8K video streaming, virtual and augmented reality (AR/VR), and industrial applications. Mm Wave’s ability to reduce latency to milliseconds (MS) is a game-changer for critical services such as remote surgery, autonomous vehicles, and real-time robot control, ensuring a near-instantaneous response between devices and the network. The challenges, such as the shorter range and susceptibility to obstructions, are opportunities for innovation, requiring a higher density of antennas and infrastructure. m-MIMO (massive multiple input, multiple output) is a key player in this innovation, using multiple antennas in a base station to transmit and receive multiple signals simultaneously, significantly increasing spectral efficiency and network capacity. Its main benefits are increased capacity, improved coverage, Beamforming, and energy efficiency. M-MIMO allows multiple data streams to be sent and received simultaneously, increasing the amount of data that can be transmitted without increasing the required bandwidth. This is essential in high-density areas such as stadiums, city centres, and events. Using multiple antennas improves signal coverage, especially in urban environments with many physical barriers. This allows 5G to offer more reliable service in densely populated areas. Beamforming is the "Beamforming" technique used in m-MIMO that directs signals directly to devices instead of scattering them in all directions. This improves connection quality and reduces interference, which is essential for critical applications requiring high reliability, such as remote surgery and precision robotics. By optimizing the direction and use of spectrum, m-MIMO can reduce the power consumption of networks, an essential consideration for the sustainability of large-scale telecommunications networks (Alessandro et al., 2024).

In contrast, AR enhances medical training by overlaying digital images, such as detailed anatomy or instructions, on top of the physical environment. This allows medical professionals to learn by interacting with 3D models directly in the hospital or operating room environment. For example, doctors can use AR to visualize the inner layers of the human body while performing dissections or surgeries on mannequins.AR allows surgeons to plan procedures more accurately. Using preoperative images (such as CT scans and MRIs); surgeons can design 3D models of a patient's organs and systems. Surgeons can visualize these models superimposed on the patient's natural body during the operation. This is particularly useful in complex surgeries, such as neurological or orthopaedic surgeries, where every millimetre matters. AR also maps tumours or bone structures, helping surgeons plan minimally invasive surgeries more accurately and safely.AR has been already used in live surgeries to provide surgeons with real-time visual assistance (Dhar et al., 2021).

For example, using AR glasses, a surgeon can see detailed patient information, such as CT or MRI images, superimposed directly on the surgical field. This helps the surgeon make quick and accurate decisions. One example is the use of AR in orthopaedic procedures, where prosthetic cutting guides has projected for the patient's body to ensure the prosthetic fits accurately. VR proceeds for physical and cognitive rehabilitation. Patients who have suffered injuries, strokes or surgeries can use VR to practice movements and activities within a controlled virtual environment. This helps speed up the recovery process and improves patient engagement. For example, patients with motor difficulties can use VR programs to practice arm or leg movements, immersing them in games or simulations that stimulate the brain to relearn specific movement patterns. AR operates to help patients perform rehabilitation exercises at home, with visual guides on their physical routines. The system provides real-time feedback, helping them improve their exercise performance. VR works in the treatment of phobias, anxiety disorders and post-traumatic stress disorder (PTSD). Through controlled simulations, therapists can expose patients to feared situations safely and progressively, helping them overcome their phobias. For example, patients with a fear of heights can be exposed to simulated height situations in a virtual environment until they feel comfortable (Kassutto et al., 2021; Matthews & Shields, 2021).

# 2. Conceptual Background

## 2.1. Technological Innovation System (TIS) Framework

Carlsson et al. (2002), aligned with Freeman & Perez (1988), have suggested that an Innovation System is a systemic perspective in order to a conjunction of actors, institutions, networks and technologies with the same objective of directing and intensifying the speed of innovation (Carlsson et al., 2002; Markard, 2020). In other words, the systems perspective, assumed from an engineering viewpoint, comprises components, attributes, and relationships that link all these concepts in an interactive approach. On the other hand, technology is not a component of TIS but precisely what composes all these systems. The importance is in the framework for the substantial relations of the Innovation process, regarding an economic or a techno-economic perspective of organisational, integrative or coordinated ability. Hekkert et al. (2007), this approach in TIS terms may direct the speed and intensity of innovation, proposing "indicators" for the Innovation Policies respecting a specific technological system, whatever it is (Hekkert et al., 2007).

Furtado et al. (2020), Silva de Oliveira et al. (2022), Ashari et al. (2022), Bergek (2015) and Markard et al. (2015) show the conceptual background fragility of the TIS is related to geographical contexts (Markard, 2020; Edsand, 2019). Part of the worldwide literature has tried to resolve (Van Der Loos et al., 2021; Walrave & Raven, 2016; Huang et al., 2016 and Edsand, 2019, for example) this problem with the micro and meso spheres in its analytical perspectives, which means regionally and nationally, but not always in a multilevel perspective (MLP). Binz and Truffer (2017, 2011) and Bergek, Hekkert, Jacobson, Markard, Sandén and Truffer (Bergek et al., 2015) are crucial papers for comprehending the problem of contexts in Technological Innovation Systems, especially in connection to decision-making and political perspectives in a policy mixed approach.

Contrastingly, exponents such as Knut Koschatzky, Roald A. A. Suurs and Knut Blind adopt a lens that centres more decisively on the structural dynamics and developmental pathways ingrained within innovation systems. Their focus lies on structural *motors* or *drivers* that shape TIS over time, delineating trajectories that technological evolution traverses. Beyond structural modifications, their purview encompasses intricate metamorphosis and reconfiguration that TIS undergoes over technological trajectories. Via technological paths, TIS grasps technologies' directional evolution and maturation of the technological life cycle, explaining the mechanisms that culminate in certain technologies asserting dominion within the TIS framework. Consequently, they underscore the pertinence of *disruptive innovations*, heralding transformative shifts in the innovation industrial configuration (Ashari et al., 2023; Blind & Niebel, 2022; Buggenhagen & Blind, 2022). One idea of the present paper is to integrate all these elements of the Technological Innovation System discussion and perspectives.

# 3. Material and Methods

## 3.1. Methods

Hekkert et al. (2007), Furtado et al. (2020), Johnson & Jacobson (2001), Negro (2009), and Binz & Truffer (2017) have each added unique contributions to this framework, emphasising the profound influence of TIS functions in guiding the trajectory of technological change. Conversely, the perspectives achieved by Bergek et al. (2015) and Weber & Truffer (2017) underscore the significance of TIS mechanisms, especially those related to blocking, in shaping innovation contours. In this sense, this investigation proposes working with ***TIS Functions 2 and 3***. These functions can sign precisely how technology transfer and spillover have produced concerning 5G technology. This structural focus of analysis can be understood in the figure below:

**Figure 1 - TIS Working Functions (See Hekkert et al., 2007; Furtado et al., 2020; Martinelli & Mazoni, 2024; Martinelli et al., 2024)**

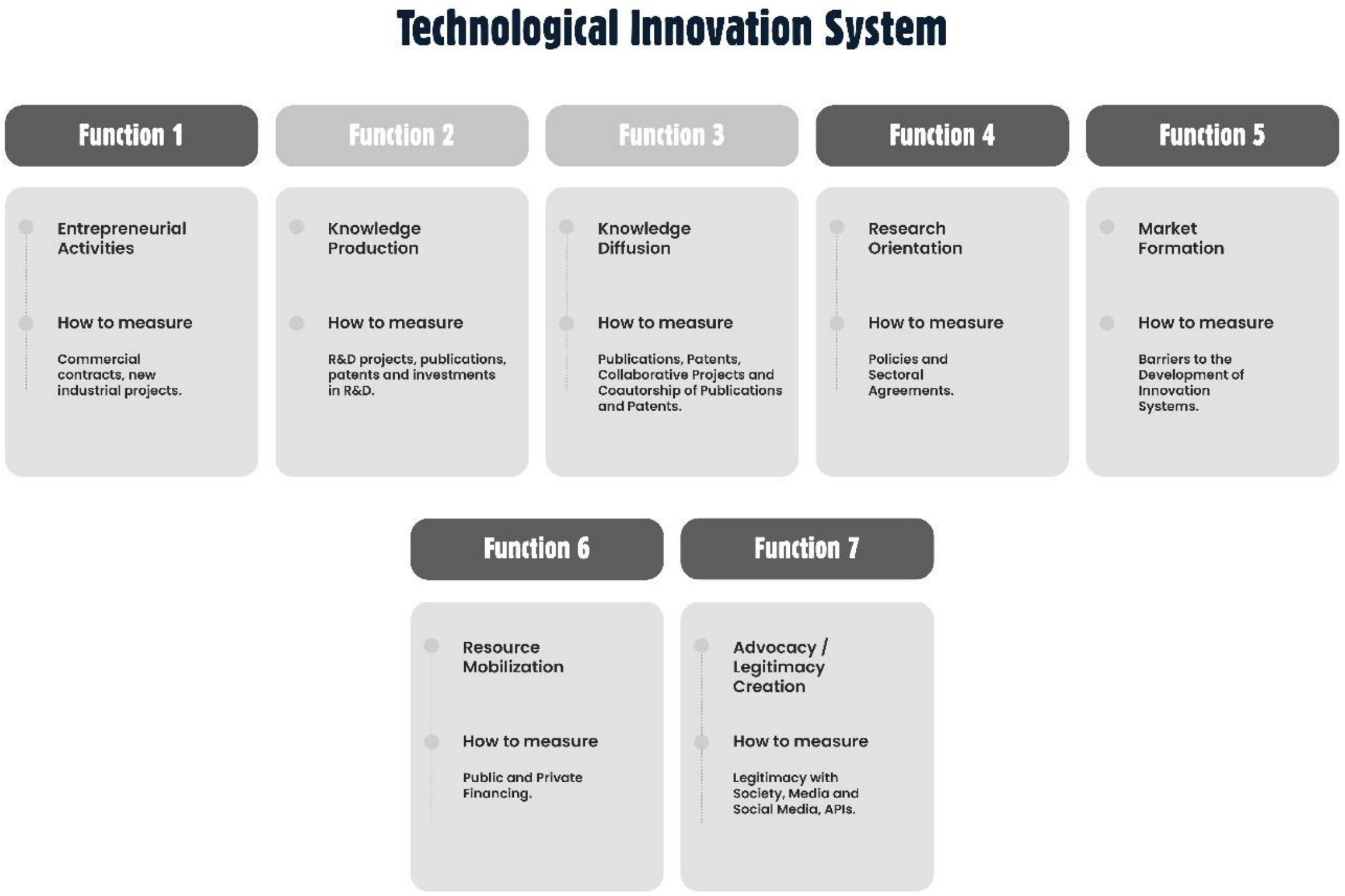

Henceforth, this article aims on publications, patents and network scientific collaborations, which have, for instance, a crucial role in the speed and direction of the innovation process. Publications and patents have relevant elements that reinforce 5G and 6G development (Buggenhagen & Blind, 2022) and sustain efforts in medical applications globally. In telecommunications, R&D has been established as a cornerstone of innovation, diffusing via channels of *knowledge* and *technology transfer* encompassing mainly ***publications and patents*** (Ashari et al., 2023; OECD & Eurostat, 2018). However, it has also been established that more than a diffusion of publications and patents, R&D is an indicator of publications and patents in terms of results. These endeavours have amplified the possibilities of research becoming an accepted application in the market. Nevertheless, recently, there has been a long journey for technologies to gain reliability in the market (Ashari et al., 2023, Markard, 2020).

The present study is a "***Lundvallian***" (Lundvall, 2016) analysis when our study focuses on the Learning Economy' approach, studying only the ***Knowledge Functions***. On the other hand, it is Rickap & Lundvall' (2021) perspective when interplaying the Technological Innovation System (TIS) framework with the geographical contexts (Edsand, 1996; Rickap & Lundvall, 2021) in a National TIS approach (Schmidt & Dabur; Binz & Truffer, 2017).

## 3.2. Empirical Case of 5G/6G for Remote Surgery

This study has been based on a theoretical-empirical case focused on Remote Surgery. The Healthcare TIS has interconnected with the 5G TIS and 6G TIS through publications, patents and networks, elucidated by research on clustering and collaborations. In this way, the case of Remote Surgery empirically retraces the conceptual framework of TIS through its primary functions for a nascent technology: knowledge production and network formation.

This would serve as a foundation for the dynamics of technological overflow - *spillover* - and technology transfer, based on the knowledge produced in publications and patents terms and facilitated by the formation of networks. Innovation cycles rely on these technological changes and feed on these learning curves, enabling innovation actors in 5G and 6G to adjust their route corrections (Rodríguez-Pose & Crescenzi, 2008; Preobrazhenskiy & Firsova, 2020; Lundvall, 2016; Rikap & Lundvall, 2021; Ashari et al., 2023).

Driving of institutions in 5G and 6G in interconnection with the healthcare system, given its best industrial application, the Medical Internet of Things (medical-IoT, m-health, IoMT), has the best way to be patentable and to build arguments in article publications in Remote Surgery. It maps entrepreneur activity (Dwivedi et al., 2022; UNCTAD, 2021).

This idea refers to the role that institutions, such as governments, universities, technology companies, and regulatory bodies, play in implementing and developing 5G and 6G technologies, with a focus on their application in the healthcare system. This involves creating policies, regulations, investments, and strategic partnerships to ensure that the infrastructure and services of these new networks can be used effectively in the healthcare sector. The interconnection between these networks and the healthcare system can enable advances in areas such as telemedicine, remote surgery, real-time patient monitoring, and more accurate diagnostic systems, as well as improve the quality of medical services and accessibility, especially in remote areas. It is about how these institutions drive innovation and integrate new connectivity technologies to transform healthcare delivery.

**Figure 2 - TIS Levels and Working Functions (See Schmidt & Dabur, 2014, and Binz & Truffer, 2017)**

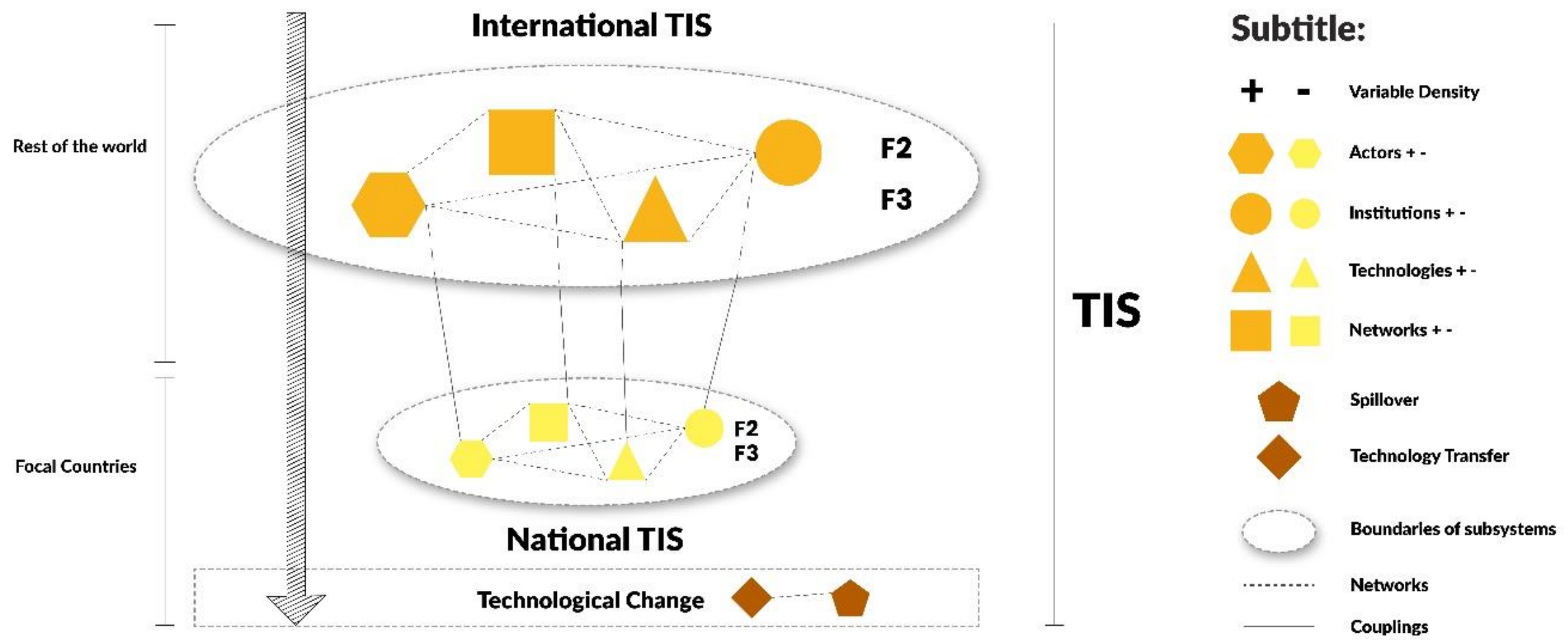

Figure 3 outlines two main steps in the analysis at the international and national levels. This approach, applicable to any geographical context, is suitable for any study case in 5G networks. For instance, when examining a national geographic context, such as Germany, we can determine the intensity and speed of innovation through regional data collected at the national level. This approach can help answer questions about technology transfer and technological spillover. In the context of Figure 3, it is important to highlight the role of the national TIS in international cooperation, as it can attract technological foreign direct investments. At the national level, Average Citation Impact has also been utilised to compare the number of patents by affiliation to demand-pull understanding. The interaction between International TIS and National TIS makes technological change possible in global TIS and National TIS, interplaying both systems.

The empirical results of specific industrial applications research, such as Medical Industrial Internet of Things (MIIoT) in remote surgery cases, make possible the route correction of the TIS context problem proposed by the current literature (Bergek et al., 2015; Furtado et al., 2020; Silva de Oliveira et al., 2022; Edsand, 2019). According to publications and patents, using the industrial application to the national contexts suggests macro and meso-economic contexts, which may complement the TIS conceptual background in a more specific approach than the worldwide context (Schmidt & Dabur, 2014; Binz & Truffer, 2017)[1]. After all, when we work with patents, it is crucial to highlight meso-level aspects for application efforts. Due to the rigid technological niche distinctions, the present study has comprised the international and national TIS perspectives, not in a multilevel approach.

[1] See Schmidt & Dabur's status in item 2.4.1.

# 4. Results of Knowledge Production and Diffusion Analysis

## 4.1. Sample I: Knowledge Production (Function 2) Quantifying

**Table 1 -Publications and Patents for 5G/6G and remote Surgery (Web of Science, 2024, and Orbit Intelligence, 2024)**

| **PUBLICATIONS (2016 - 2024)** | **PATENTS (2016 - 2024)** |
| --- | --- |
| International TIS – 160 papers | International TIS – 5,704 patent families |
| National TIS – 5 papers in Germany | National TIS – 51 patent inventions in Germany |

Especially in WoS, the number of publications may vary according to the selected period. Thus, in a month, there can be a discrepancy in publications and patent terms. Crucial issues that emerged from the patents subject for 5G are the correspondence of SEPs (Standard Essential Patents) to the correlated standards for 5G and then for 6G, according to the international debate. Some authors (Buggenhagen & Blind, 2022) raised the subject; however, they do not have an in-depth viewpoint focusing on the 5G and 6G Releases and it also only works with these applications, so more studies are still necessary on this subject.

**Graph 1 - Publications and Patents for 5G/6G and remote Surger by Timeframe**

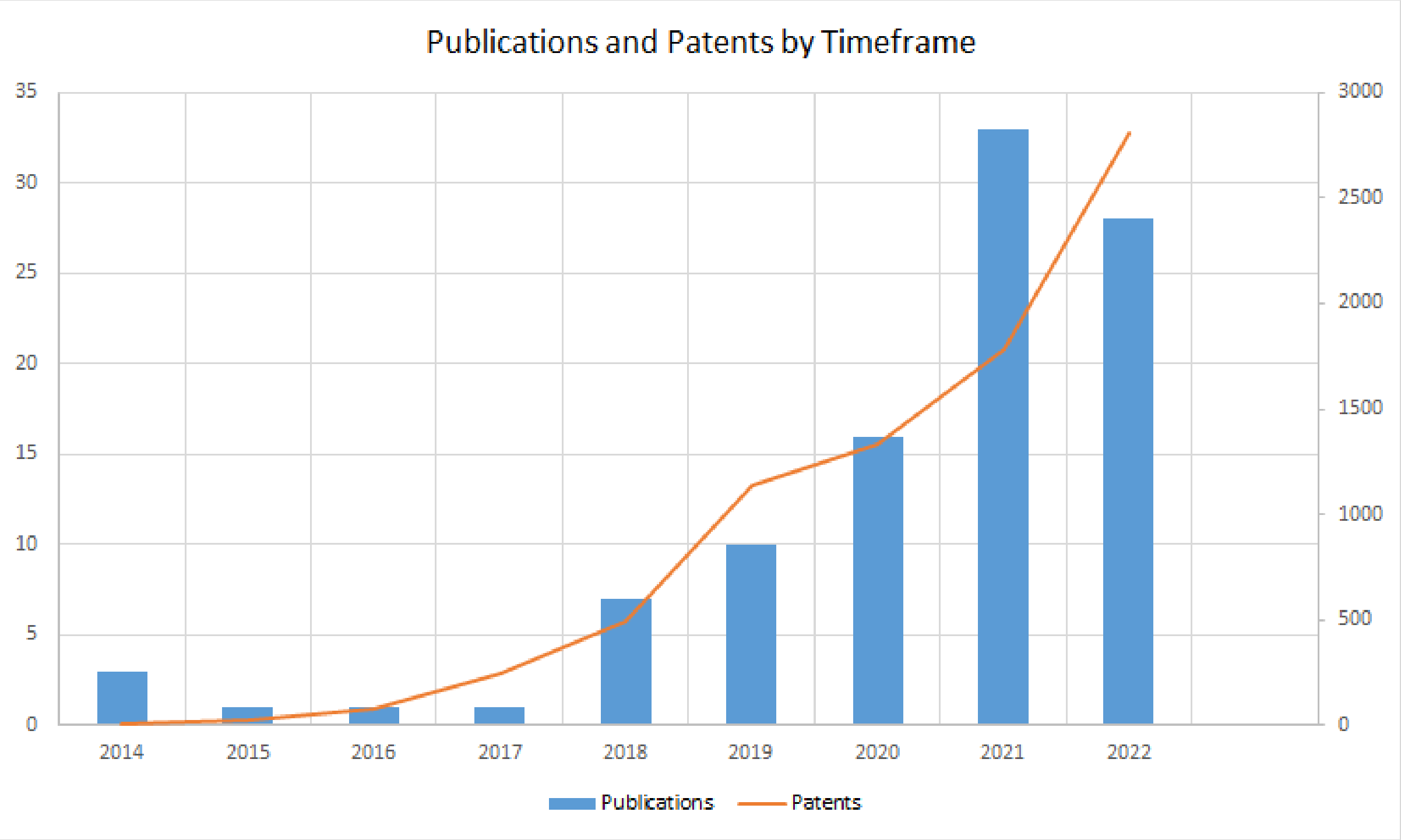

Graph 1, which outlines inferred patents in 5G or 6G and Remote Surgery, illustrates a notable trend beginning in 2014. The activity in these areas markedly increased after 2018, with a significant surge in 2019 in terms of patents. This pattern can be closely linked to several pivotal events and developments in the telecommunications industry, particularly the spectrum auctions in South Korea, North America, and Brazil. The early 2000s marked the inception of research and development in 5G, 6G networks and remote surgery applications. Foundational patents have filed during this period, setting the stage for future advancements. After 2014, there was a noticeable uptick in patent filings, reflecting intensified R&D efforts. The year of 2019 stands out with a substantial peak in patent filings, a testament to the culmination of years of research, the ramping up of 5G deployments, and the significant role of spectrum auctions in driving technological advancements in this competitive field.

South Korea has been a frontrunner in adopting and deploying 5G networks. The countries early and extensive spectrum auctions facilitated rapid 5G rollouts, encouraging innovation and patenting in this domain. The United States, with its significant auctions for the 5G spectrum, has been a primary driver of global 5G innovations. These auctions provided the bandwidth for 5G networks and spurred a wave of related technological innovations, including in remote surgery applications. In 2021, Brazil's 5G spectrum auction, one of the largest in Latin America, opened up significant opportunities for domestic and international players. The auction's impact is evident in the increased patent activity as companies positioned themselves to capitalise on the new market dynamics. The flurry of patent activity emphasises the dynamism and energy in the industry, with enterprises and research institutions racing to develop new technologies, secure intellectual property, and gain competitive advantages in the field. The patent trends show the strategic significance of spectrum auctions, as they are not merely administrative processes but catalysts for technological advancements and market positioning. Data highlights the intense global competition in 5G and 6G networks. Notably, regions that have conducted successful spectrum auctions, such as South Korea, North America, and Brazil are emerging as key players in the international TIS. We see the rise of these economies because of a combination of strategic factors, particularly concerning connectivity infrastructure. The phrase suggests that the global competition around 5G and 6G networks is shaping a new economic landscape, and regions that have successfully auctioned spectrum, such as South Korea, North America and Brazil, are positioning themselves as essential players in the International Technology Innovation System (TIS).

These countries have been proactive in implementing new technologies, which drives domestic development and puts them at an advantage in the global race to lead telecommunications innovation. The ability to take advantage of 5G and 6G networks is directly linked to economic success, and these regions are emerging as examples of how well structured policies and strategic investments can generate competitive advantage. Their strategic spectrum auctions have significantly shaped the TIS. The surge in patent activity post-2010, peaking in 2020, reflects the culmination of extensive R&D efforts and the strategic importance of these auctions. As these regions continue to lead in 5G deployments, they will likely influence global trends and drive further innovation in next-generation telecommunications and related fields.

**Graph 2 - Publications and Patents for 5G/6G and remote Surgery by Affiliations/Organisations**

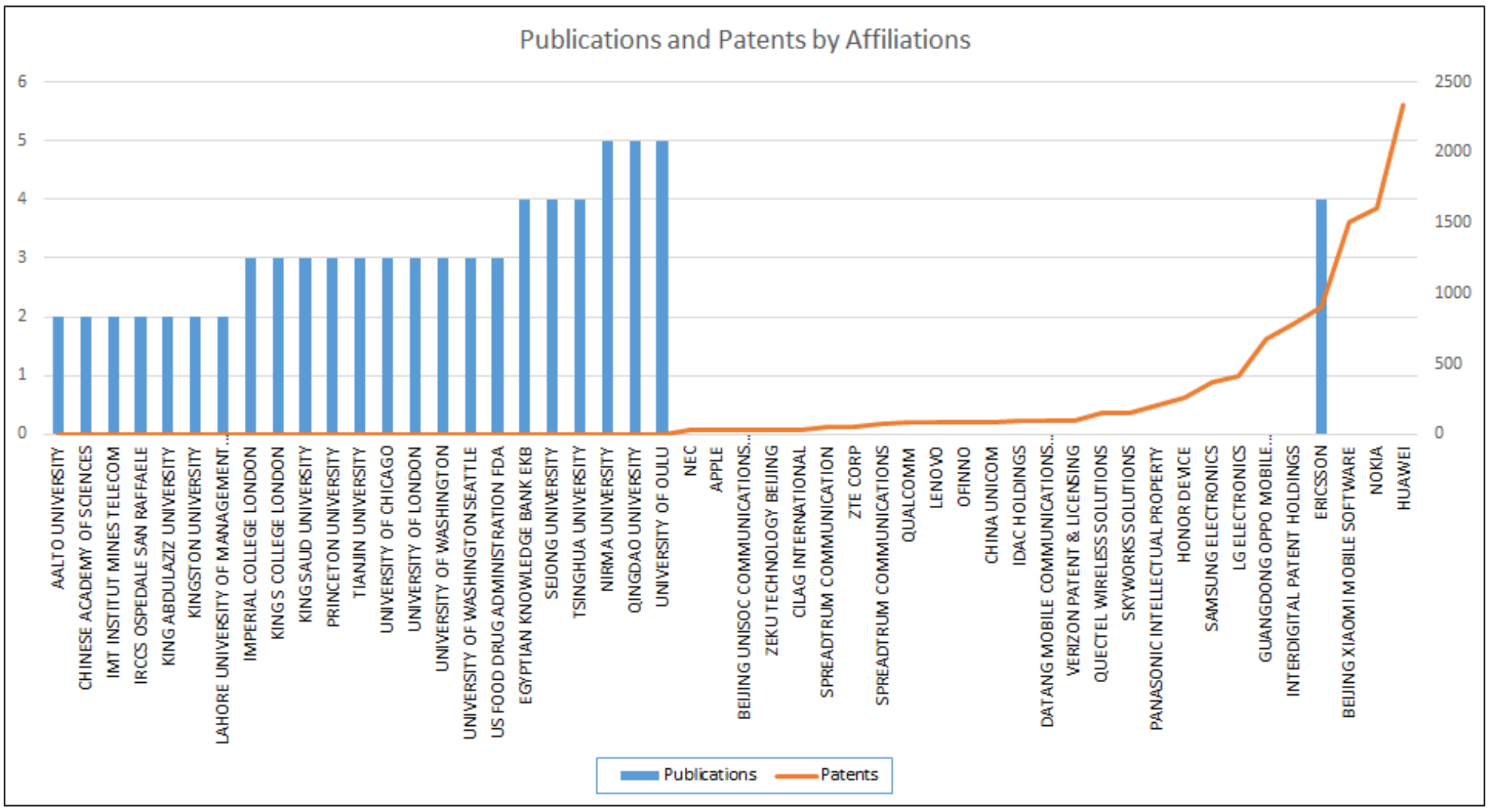

Graph 2 compares the number of patent families and academic publications produced by various institutions and companies. The blue bars represent the count of patent families, while the orange line indicates the number of academic publications. It can be observed that a high level of patent activity is observed; Huawei, Nokia, Beijing Xiaomi Mobile Software and Ericsson show significant numbers of patent families, suggesting a strong emphasis on intellectual property development. Second, Institutions like the University of Oulu, Qingdao University, and Nirma University have higher numbers of academic publications, indicating their active involvement in research and dissemination of knowledge. Third, LG Electronics, Samsung Electronics, and Samsung Electronics exhibit notable figures in patent families and publications, reflecting a balanced approach to innovation and academic research. This balance reassures us of the industry's commitment to both innovation and research. In addition, some institutions, such as US Food Drug Administration FDA and University of Washington Seattle show minimal patent activity but have a presence in academic publications, highlighting their focus on research rather than patenting. Finally, Graph 2 indicates various organisations' diverse strategies and focal points in innovation and academic research. Overall, it is an interactive process, sometimes focusing on academic research in a science-push initiative, on the industry in a demand-pull initiative or in both of them.

**Graph 3 - Publications and Patents for 5G/6G and remote Surgery by Countries**

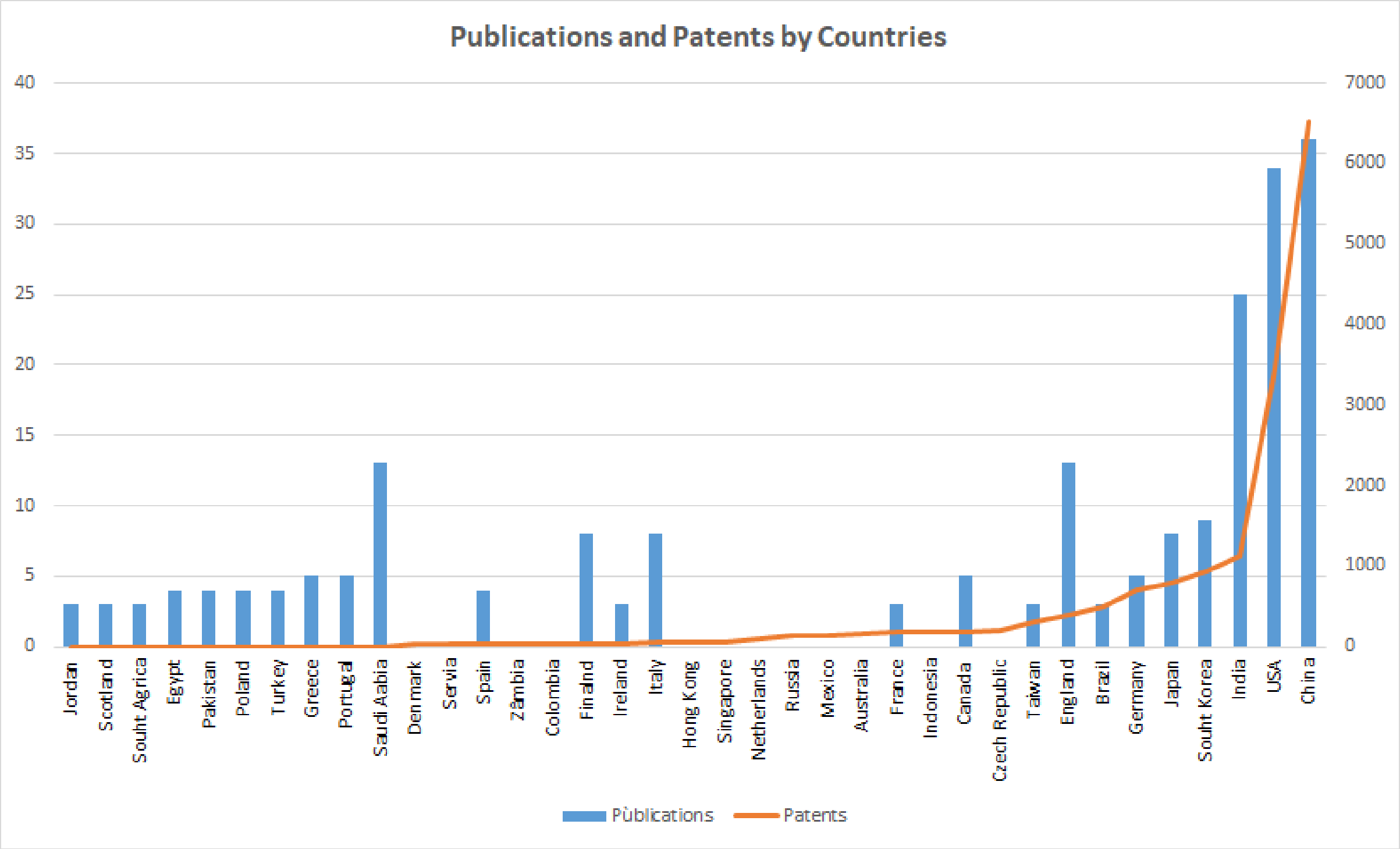

Graph 3 provides an insightful comparison of the number of academic publications and patents across various nations. The blue bars denote the count of academic publications, while the orange line represents the number of patents. China has the highest number of patents, surpassing 7000, indicating a robust emphasis on securing intellectual property. The USA follows with many patents, highlighting its intense innovation system. Countries with high academic publications, such as India and Germany, have considerable academic publications, demonstrating their active engagement in research. The USA, China, and Japan contribute significantly to academic literature. The USA and Germany exhibit a balanced approach, contributing notably to academic publications and patents, reflecting well-rounded TIS with a prominent interactive innovation process. Asian countries like China, India, Japan, and South Korea show substantial activity in patents and publications, underscoring the region's growing prominence in global research and development. Germany, France, and Finland (all European) actively participate in both areas, highlighting their established research infrastructures. Some, such as Saudi Arabia and Singapore, have a noticeable number of patents but fewer academic publications, suggesting a stronger focus on applied research and development. Those facts have crucial implications on Innovation Hubs, Research Intensity and Strategic Patenting. This is seen as the USA and China emerged as key innovation hubs with significant contributions to patents and academic publications. Moreover, countries with high academic publications, like India and Germany, play a critical role in advancing knowledge and fostering global research collaborations. In addition, the substantial patent activity in countries such as China and the USA highlights the strategic importance of intellectual property in driving economic and technological competitiveness.

## 4.2. Sample II – Knowledge Diffusion at International TIS (Function 3)

### Graph 4 - Publications on 5G/6G and Remote Surgery at International TIS by Keywords

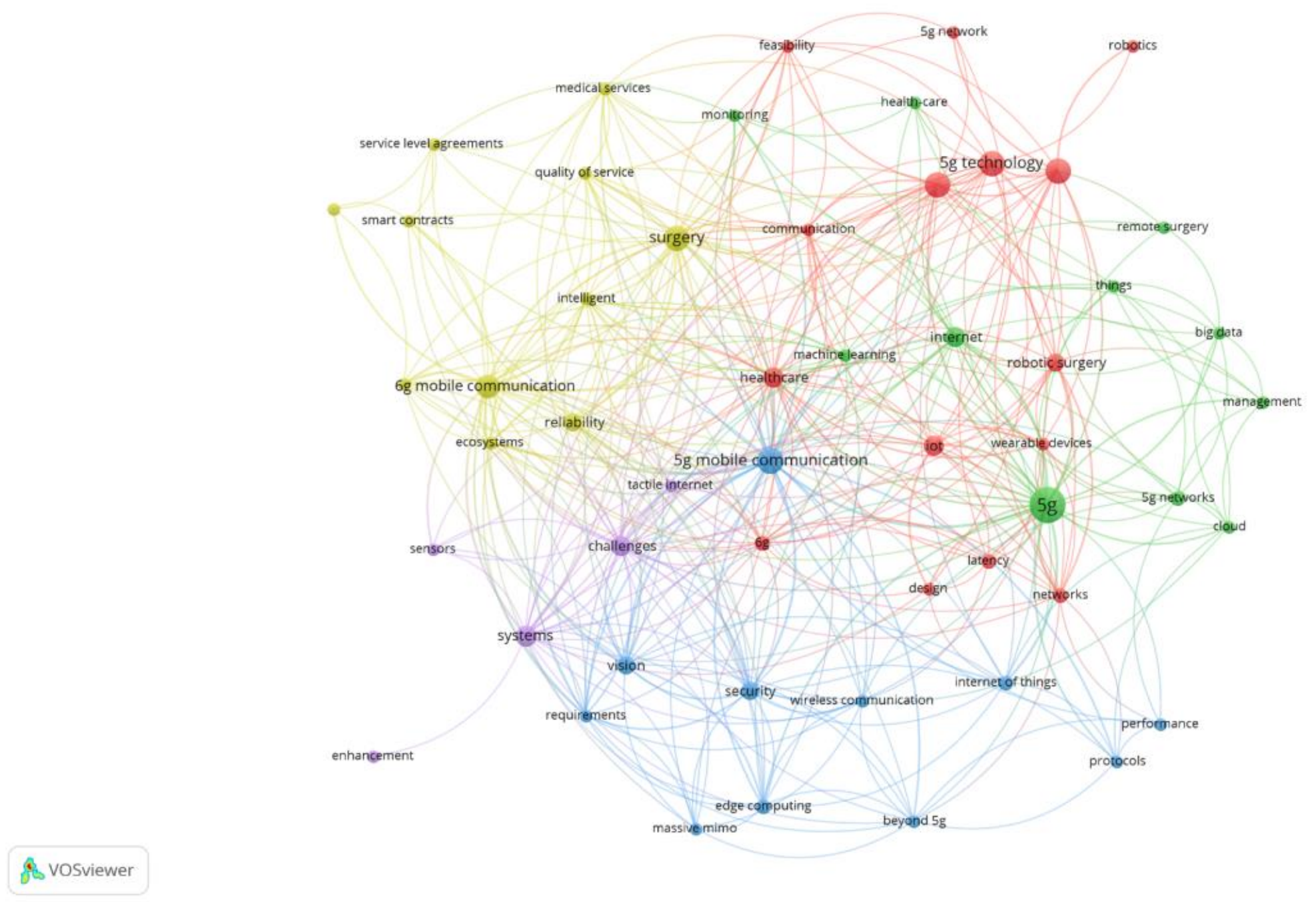

### Graph 5 - Zooming in the Surgery’s Cluster

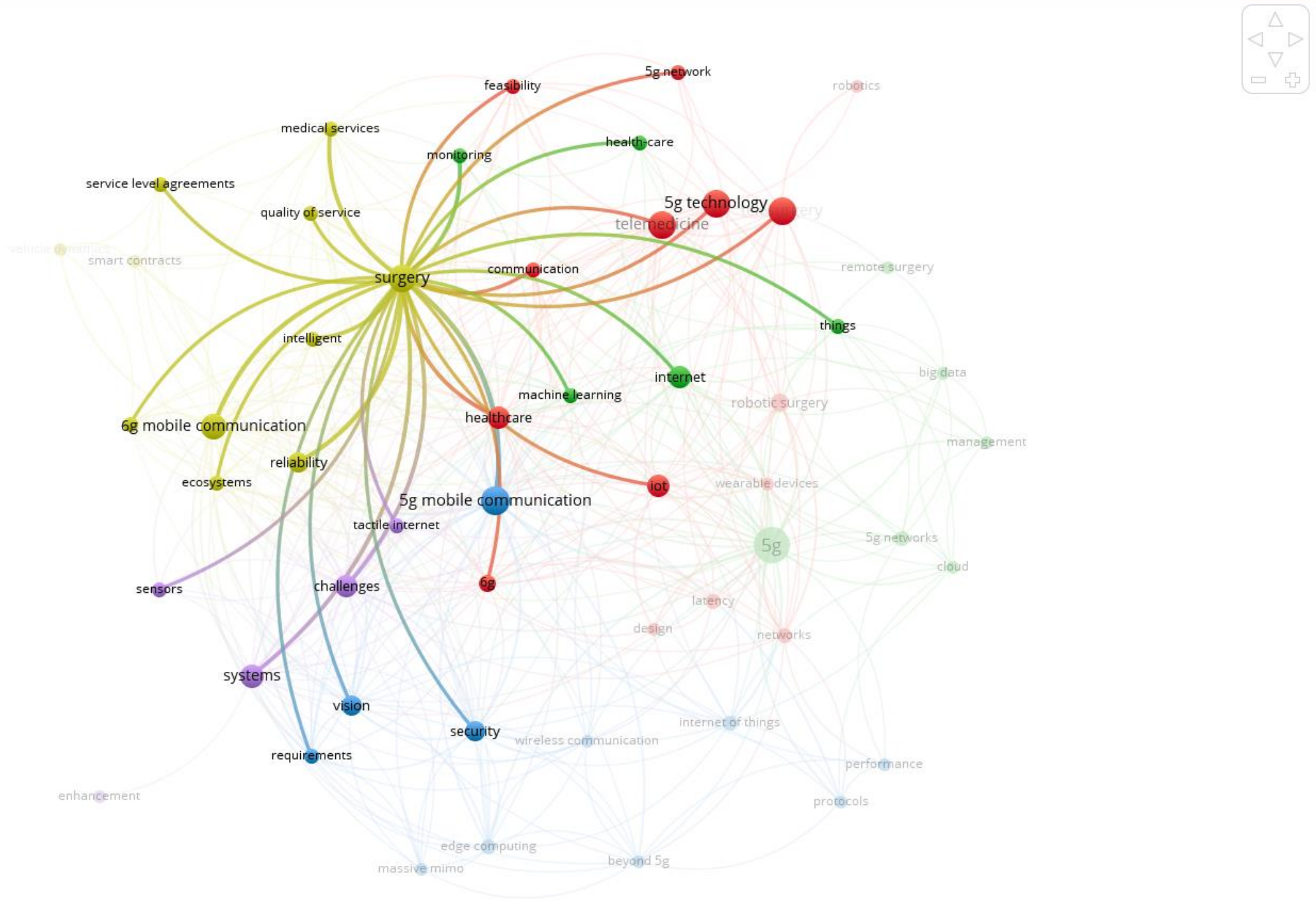

Graph 4 shows the visualisation presented is a network map created using VOSviewer that illustrates the relationships and connections between many critical terms in the fields of 5G and 6G mobile communication, as well as remote surgery and related technologies. The nodes represent different concepts, and the lines connecting them show the co-occurrences and interconnections between these concepts in academic literature. The term *5G Mobile Communication* is positioned, indicating its fundamental role and extensive connections with various other terms and concepts like the Internet of Things (IoT), healthcare, and wearable devices. The *6G Mobile Communication* is another term that has significant connections near the centre, reflecting the emerging interest and research in 6G technologies. The node *5G Technology* is connected to terms such as *remote surgery*, healthcare, and IoT, underscoring its pivotal role in advancing these fields. The terms 'surgery' and *Healthcare* are linked with *5G* and *6G networks*, indicating the critical applications of these communication technologies in medical services and remote healthcare delivery. Red Cluster includes terms like *5G technology*, *remote surgery*, and *healthcare*, highlighting the intersection of advanced communication technologies with medical applications. Green Cluster includes terms such as *IoT*, *big data*, and *management*, reflecting the integration of *5G technologies* with broader technological systems and *data management strategies*. Yellow Cluster includes terms like *6G mobile communication*, *intelligent systems*, and *quality of service*, indicating the focus on *future advancements* and the improvement of *communication reliability* and *performance*. However, what is crucial to note is the terms *Smart Contracts* and *Service Level Agreements*. These highlight the importance of reliable and secure communication networks in facilitating advanced digital services and agreements, shaping their necessity and value. The implications of those clustering are diverse and prominent. The network map emphasises the integrative role of 5G and 6G networks in many fields, especially in healthcare and IoT, indicating a multidisciplinary approach to research and development. However, what is even more exciting is the future Research Directions involving the prominence of 6G mobile communication and related terms. Future research is likely to continue exploring the capabilities and applications of next-generation communication technologies, opening up new possibilities and potential.
Regarding TIS, the connections between terms like smart contracts, service level agreements, and technological systems analyses the growing complexity and interdependence of modern technological ecosystems. The advanced communication networks that facilitate these connections drive this evolution. This network visualisation offers a comprehensive overview of the interconnected TIS of 5G, 6G mobile communication technologies, and their applications. It highlights the pivotal role of these technologies in various domains, particularly in healthcare and IoT, and points to the future directions and emerging trends in this rapidly evolving field.

**Graph 6 - Publications on 5G/6G and Remote Surgery at International TIS by Countries**

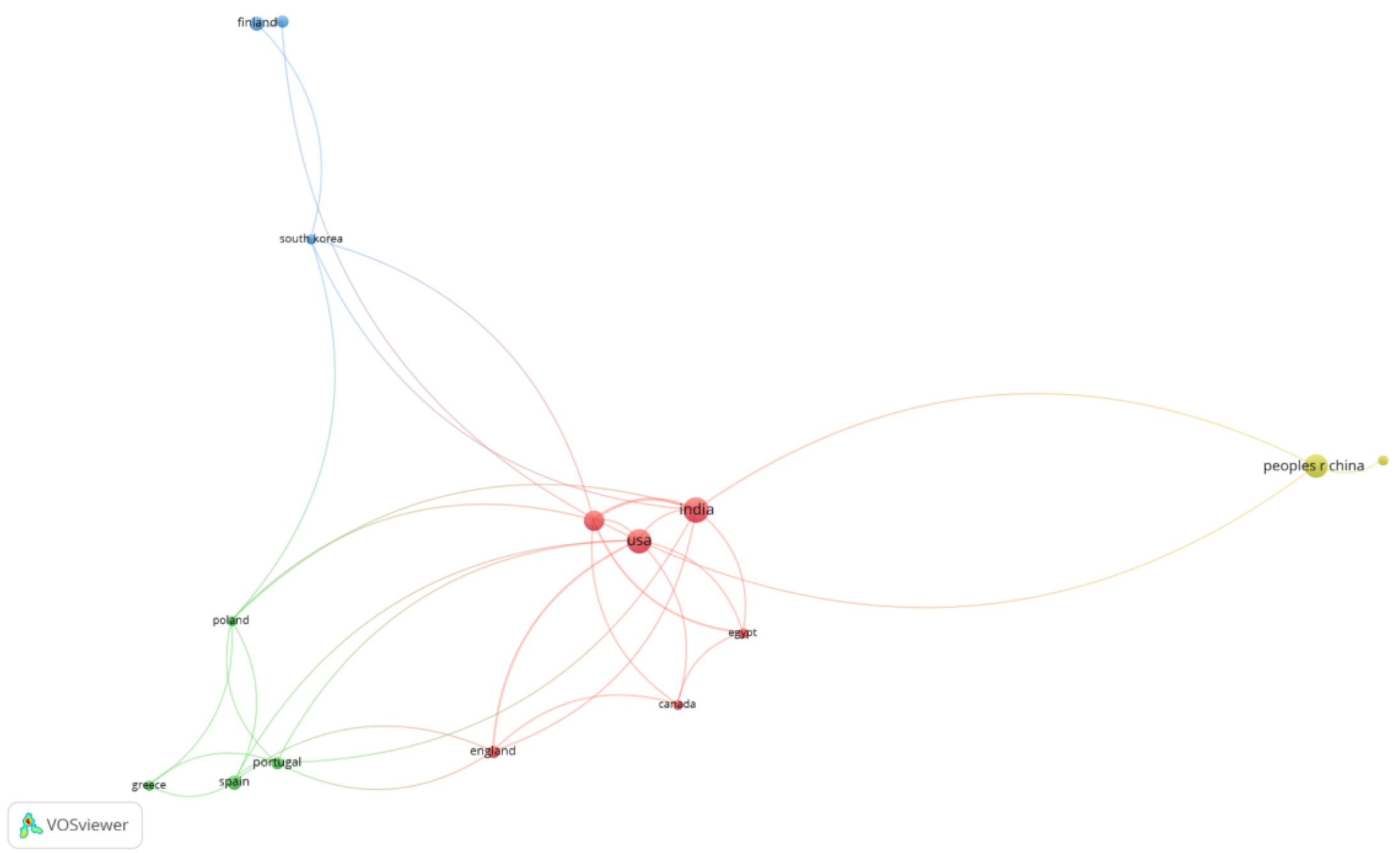

Graphic 6 provides a comprehensive indicator of the collaboration network, revealing a complex interplay of partnerships. India, the US, Egypt, Canada, and the UK are intrinsically linked to China, which forms a significant part of this network. In addition to its collaboration with India, the US collaborates with the UK, which in turn collaborates with Portugal, Spain, Greece and Poland. South Korea, Finland, India, the US, Canada, Egypt, and the UK are also part of another collaboration. Canada, as a key contributor, also collaborates with the US and the UK, highlighting the extensive nature of these partnerships. Regardless, India is not the central focus of our analysis; it is a vital participant in publications. Therefore, the nodes on the map are India, the US and the UK. Germany does not appear in Graphic 4 because it does not have any collaborations with other countries at the international level. Germany has focused more on national research, and even though there is no national programme, it has linked to the European Union.

**Graph 7 - Publications on 5G/6G and Remote Surgery at International TIS by Organisation**

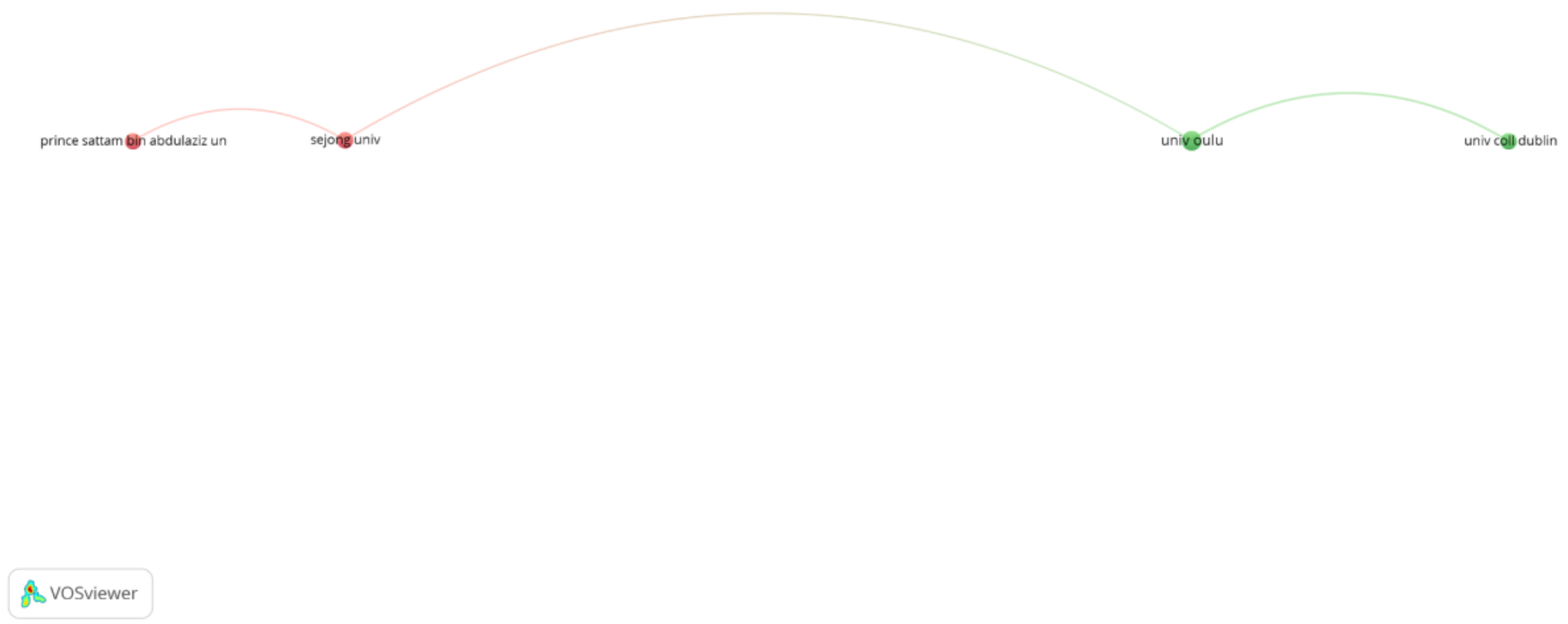

In International TIS terms, on publications, Graphic 5 highlights three main organisations: Sejong University, a private organisation located in South Korea, which preserve academic-industry relations. Sejong University has cooperation with Prince Sattam bin Abdul-Aziz University, a public organisation located in Saudi Arabia, at the same time that it has collaboration with University of Oulu, Finland. Oulu, in turn, has collaboration with University College Dublin, in Ireland.

## 4.3. Sample III - Knowledge Diffusion at National TIS (Function 2 and 3)

**Graph 8 - Publications on 5G/6G and Remote Surgery at National TIS by Countries**

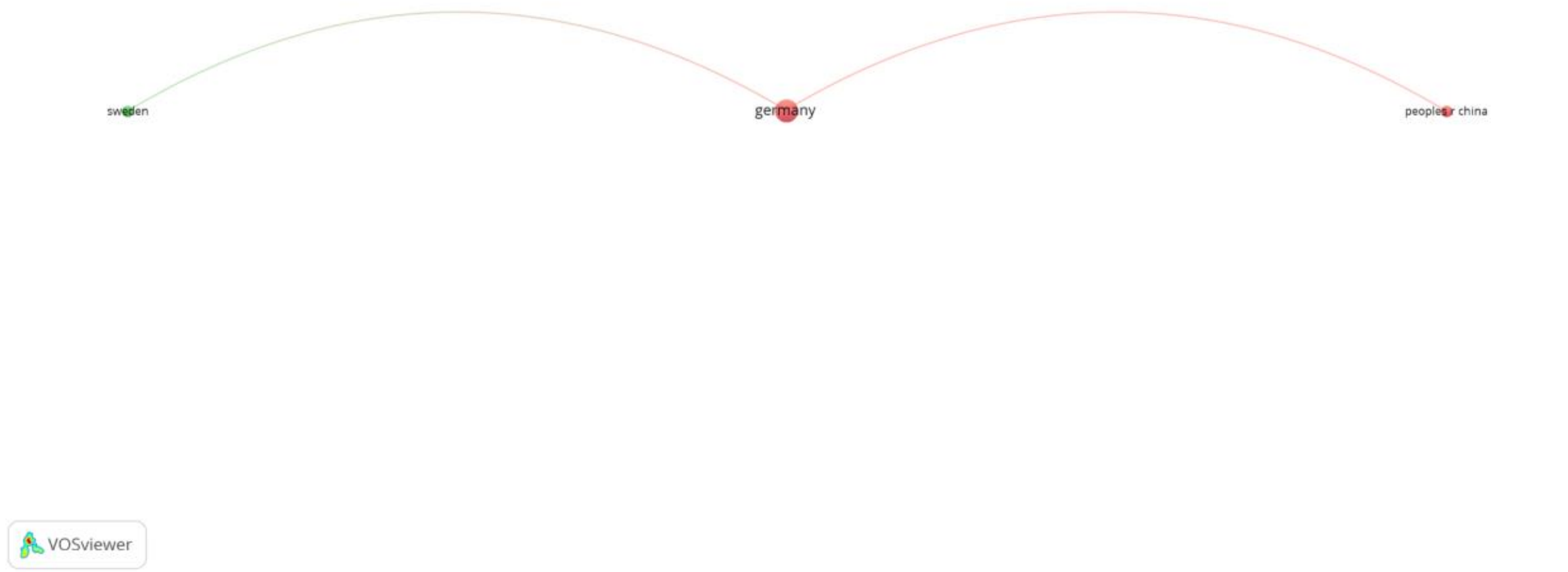

Graph 8 shows collaborations among Germany, China, and Sweden. This pattern suggests strong academic-industrial partnerships, likely driven by the market demand-pull and science-push dynamics. The involvement of leading companies such as Huawei (China) and Ericsson (Sweden) points to a shared focus on advancing technologies, possibly in the telecommunications sector, where both firms are global leaders. Germany's collaboration with these nations is a strategic alignment, leveraging scientific expertise and industrial innovation to meet market needs. It further reinforces the symbiotic relationship between academia and industry in these regions, promising a crucial future for these collaborations. On the other hand, Germany only accepts collaboration if they be in the node. On the contrary, Germany does not collaborates with other countries.

**Graph 9 - Publications on 5G/6G and Remote Surgery at National TIS by Organisations**

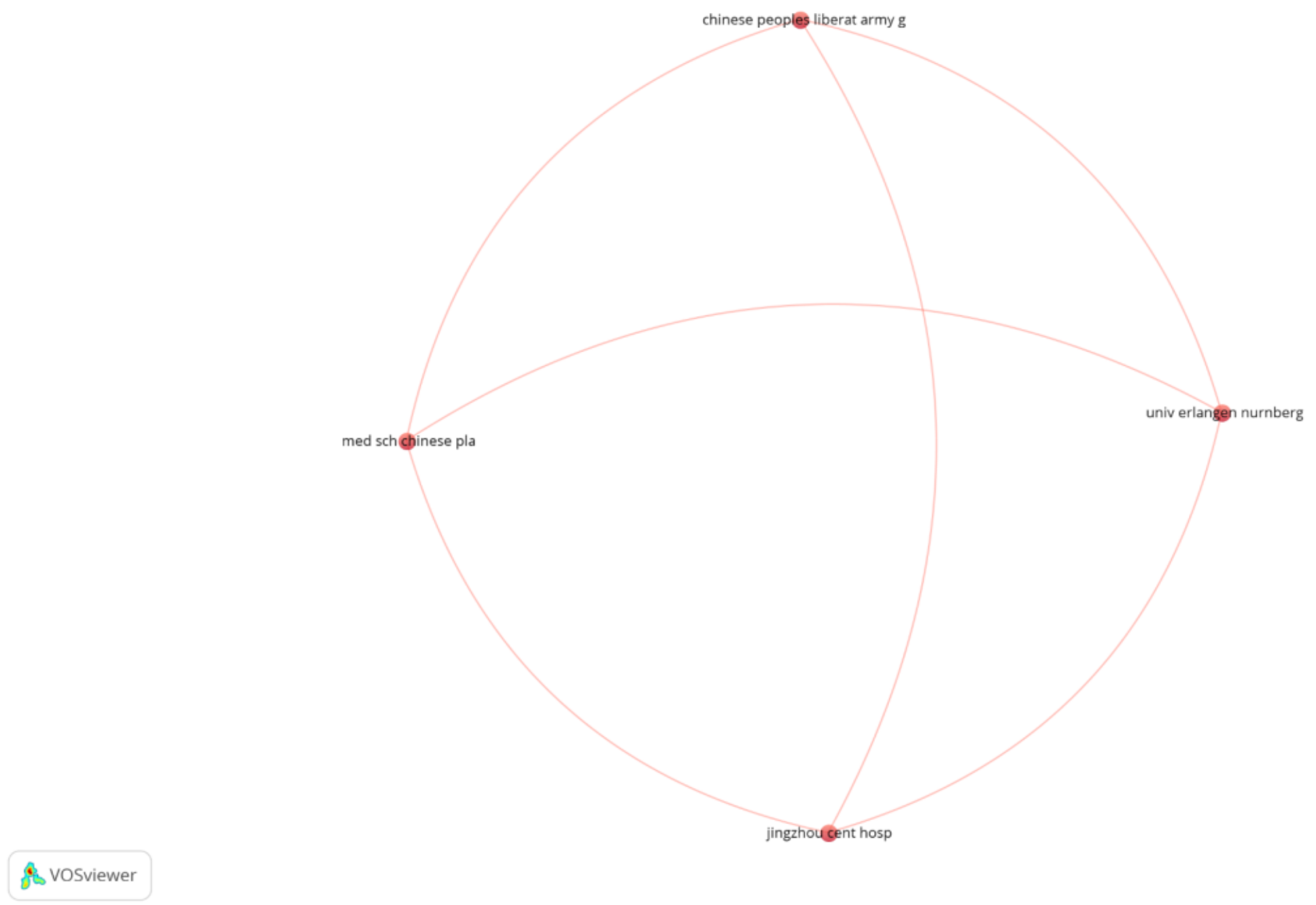

Graph 9 shows main fact that the organisation's collaboration with the Chinese People's Liberation Army General Hospital, Jingzhou Central Hospital, Medical School of Chinese PLA, and the University of Erlangen Nuremberg is more robust. Table 3 provides part of these organisations at national TIS. The chosen focus was on publications in collaboration with Germany.

**Table 2 – Eight First Organisations with a Strong Performance in German Collaborations in Publications (Web of Science and VOSviewer, 2024)**

| Organisation | Rank | Country | Publications | Citations | Total of Link Strength |
|---|---|---|---|---|---|
| Chinese People's Liberation Army General Hospital | 1 | China | 1 | 4 | 3 |
| Jingzhou Central Hospital | 2 | China | 1 | 4 | 3 |
| Medical School of Chinese PLA | 3 | China | 1 | 4 | 3 |
| University Of Erlangen Nuremberg | 4 | Germany | 1 | 4 | 3 |
| Ericsson Research | 5 | Sweden | 1 | 76 | 1 |
| Huawei Technologies Düsseldorf Gmbh | 6 | China | 1 | 6 | 1 |
| Technical University of Munich | 7 | Germany | 1 | 79 | 1 |
| Karlsruhe University of Applied Sciences | 8 | Germany | 1 | 3 | 0 |

**Graph 10 - Publications by Organisations' Citations (Web of Science, 2024)**

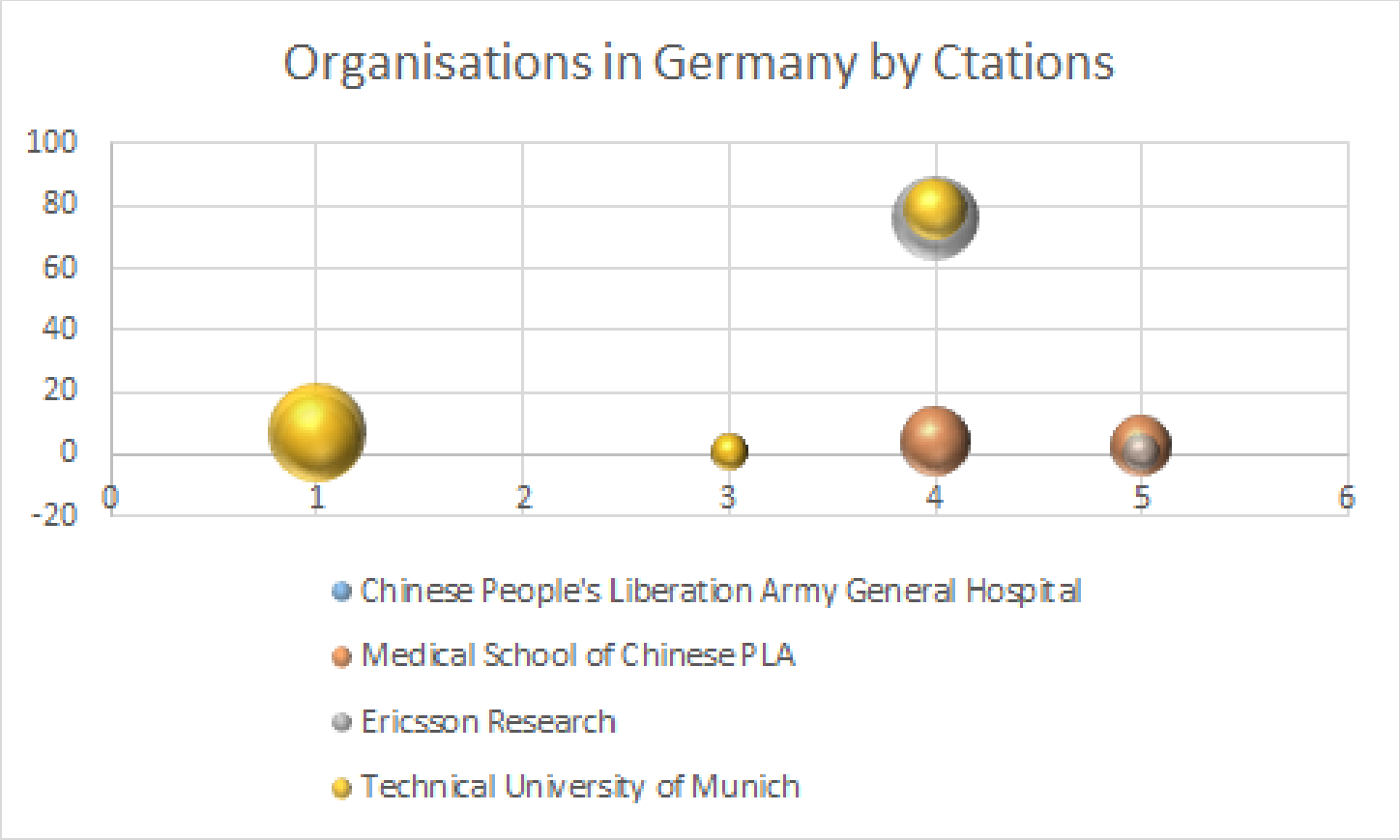

Graphic 8 inferred that Ericsson Research and Technical University of Munich have most citations on 5G/6G and remote surgery. It means that even Sweden and the United States have publications with German organisations. Ericsson, a Swedish company, has a strong presence in the US, with most of its patents deposited in the US (USPTO) and most auctions originating from the USA. As highlighted on its website (Ericsson, 2024), Ericsson plays a pivotal role in fortifying US security and leadership in 5G and beyond through strategic investments, innovative technologies, and cutting-edge services. US wireless providers, ranging from the largest to the smallest, have chosen Ericsson to cater their requirements for advanced, secure, and high-performance networks. The USA is Ericsson's most extensive global market. Since its establishment in 1902, the company has expanded to over 30 locations, including significant manufacturing, R&D, and training facilities. Recent investments of $7.3B in leading US companies further indicates Ericsson's commitment to the country (Ericsson, 2024).

## 4.4. Sample III - Patents at International TIS (Functions 2 and 3)

### Graph 11 - Patents on 5G/6G and Remote Surgery at International TIS by Protection Country (2014-2024)

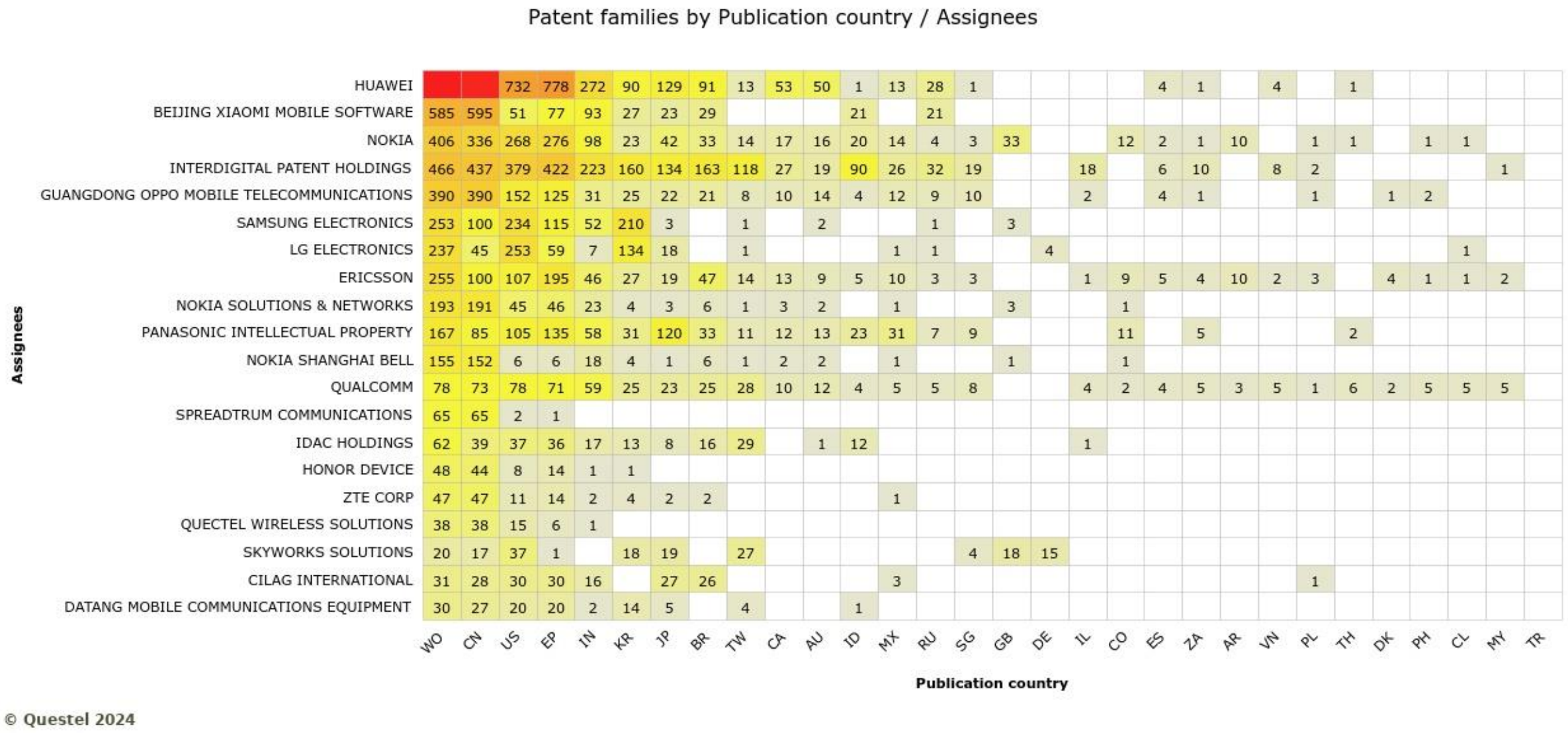

Graph 11 displays patent families categorised by publication country and assignee. The assignees on the left include major technology and telecommunications companies, while the publication countries are listed at the bottom. Huawei's dominance in the Graph is unmistakable, particularly in China (CN) and World Intellectual Property Organization (WO). This underscores their robust global and domestic patenting strategy. Beijing Xiaomi Mobile Software and Nokia also demonstrate significant patenting activity, especially in China. Xiaomi 595 patent families in China solidify its position as a major player in this market. Interdigital Patent Holdings' presence across multiple jurisdictions, with the highest activity in the US (422) and WO (437), indicates a strong focus on international protection. Oppo and Samsung's substantial patent families in China (390 for Oppo, 253 for Samsung) and WO (390 for Oppo, 253 for Samsung) reveal a balanced approach between domestic and international patents. However, it is Ericsson that stands out for its high number of patents in Europe (EP - 197) and China (255), highlighting its unwavering strategic focus on these regions. Qualcomm's significant activity in WO (78) and China (73), with additional focus in the US and Europe, further highlights the global reach of these patenting activities. Regarding this, China (CN) is a primary target for most assignees, reflecting its importance as a key market and manufacturing hub. WO (World et al. Organization) has many patent families across almost all assignees. The US (US) and Europe (EP) are critical regions for many companies, particularly Interdigital, Qualcomm, and Ericsson. CILAG International and Datang Mobile Communications Equipment have a relatively limited patent presence, focusing primarily on specific jurisdictions. CILAG, a pharmaceutical company, is investing in Advanced Medicine, which explains its limited patent presence in this context. Skyworks Solutions, on the other hand, shows activity in more diverse and specific countries like Taiwan (TW - 27), Europe (EP - 18), and Australia (AU - 18), which is less common among the other assignees. Overall, this Graph provides insight into the global patent strategies of leading technology companies, with a clear emphasis on protecting intellectual property in key markets like China, the US, and Europe.

**Graph 12 - Patents on 5G/6G and Remote Surgery at International TIS by Assignees / First Application Year**

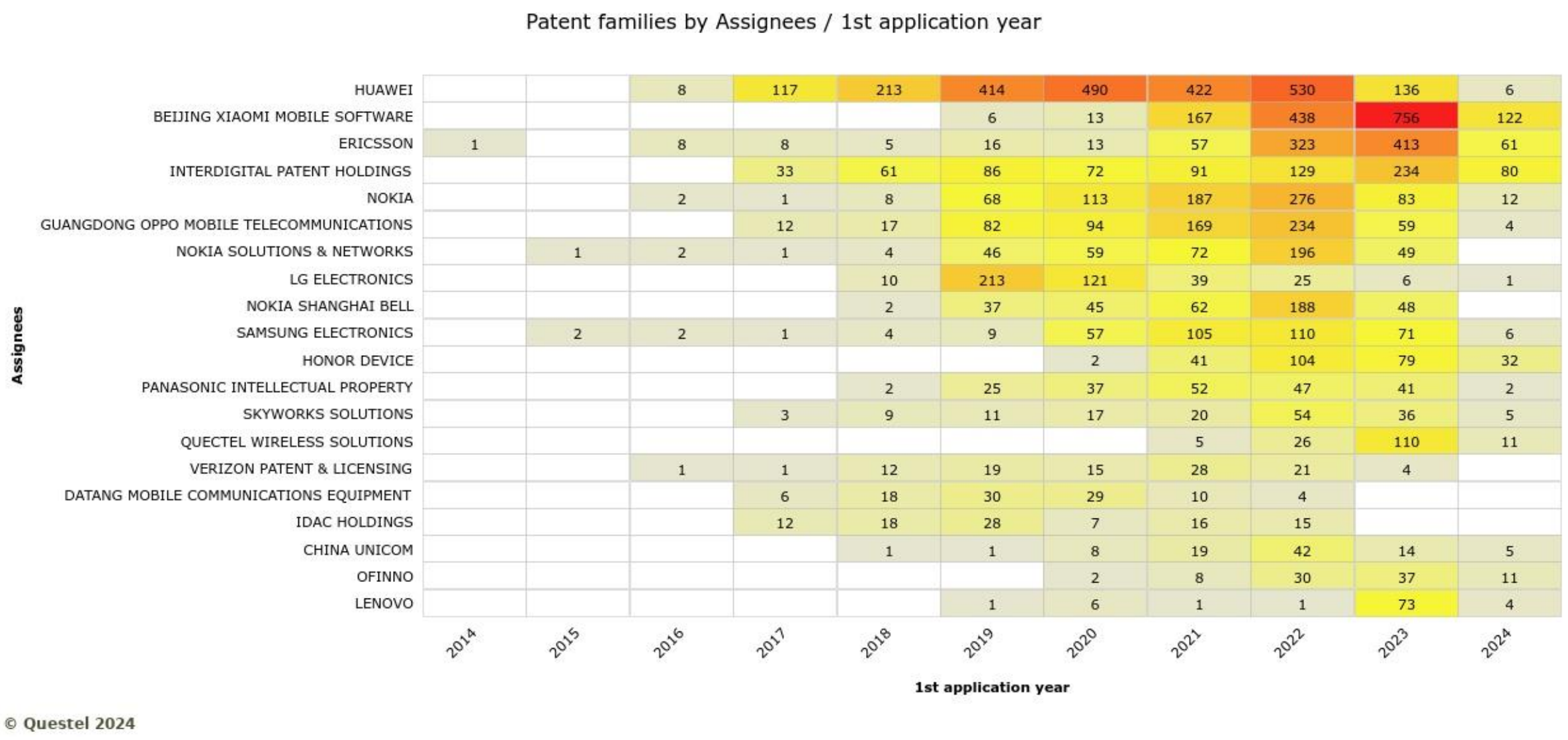

Graph 12 illustrates patent families by assignee and the year of first application from 2014 to 2024. This data displays the growth and evolution of patenting activities among key players in the tech industry. The increasing number of patent families indicates a surge in innovation and R&D investment, which can lead to the development of new technologies and products. Beijing Xiaomi Mobile Software has demonstrated an extraordinary surge in patent families, particularly from 2020 onwards. In 2021, Xiaomi had 438 patent families, a number that soared to 756 in 2022 and remained substantial in 2023 with 122 patent families. This remarkable increase not only underscores Xiaomi significant expansion and innovation in the tech space but also leaves a lasting impression of its strategy to dominate new and emerging technologies.

Huawei maintains a consistent and robust patenting strategy across the years, with notable peaks in 2018 (414), 2019 (490), and 2020 (530). Despite a decline in 2023, Huawei remains a crucial player, indicating sustained investment in research and development. This consistent strategy suggests a focus on long-term innovation and a commitment to staying at the forefront of technological advancements. Interdigital Patent Holdings and Nokia both demonstrate stable patenting activity. Interdigital has consistently spread its patent families from 2017 to 2022, with a peak of 129 patent families in 2021, a clear indication of its strong focus on long-term innovation. Nokia shows steady activity, particularly from 2019 to 2021, suggesting a targeted approach in its patent filings. Other significant players, such as Ericsson, exhibit a smaller yet stable patenting trend, with gradual increases in recent years. Oppo and LG Electronics show noticeable peaks, particularly in 2019 and 2020 for LG, highlighting their efforts in specific years to advance their patent portfolios. What truly stands out is Xiaomi patenting activity, which marks it as a standout in the industry, particularly in recent years. The company's rapid increase in patent families since 2020, with significant peaks in 2021 and 2022, underscores its aggressive strategy in capturing market share and driving technological innovation. As indicated by its prolific patent filings, Xiaomi rise suggests a focused effort on solidifying its intellectual property base, positioning itself as a leading force in the international TIS.

In summary, the Figure portrays Xiaomi as an emerging leader with rapidly growing influence, while Huawei remains a dominant player. Others, like Interdigital and Nokia, contribute steadily to the technological advancements in the industry.

**Graph 13 - Patent Collaboration by Organisations on 5G/ 6G and Remote Surgery at International TIS**

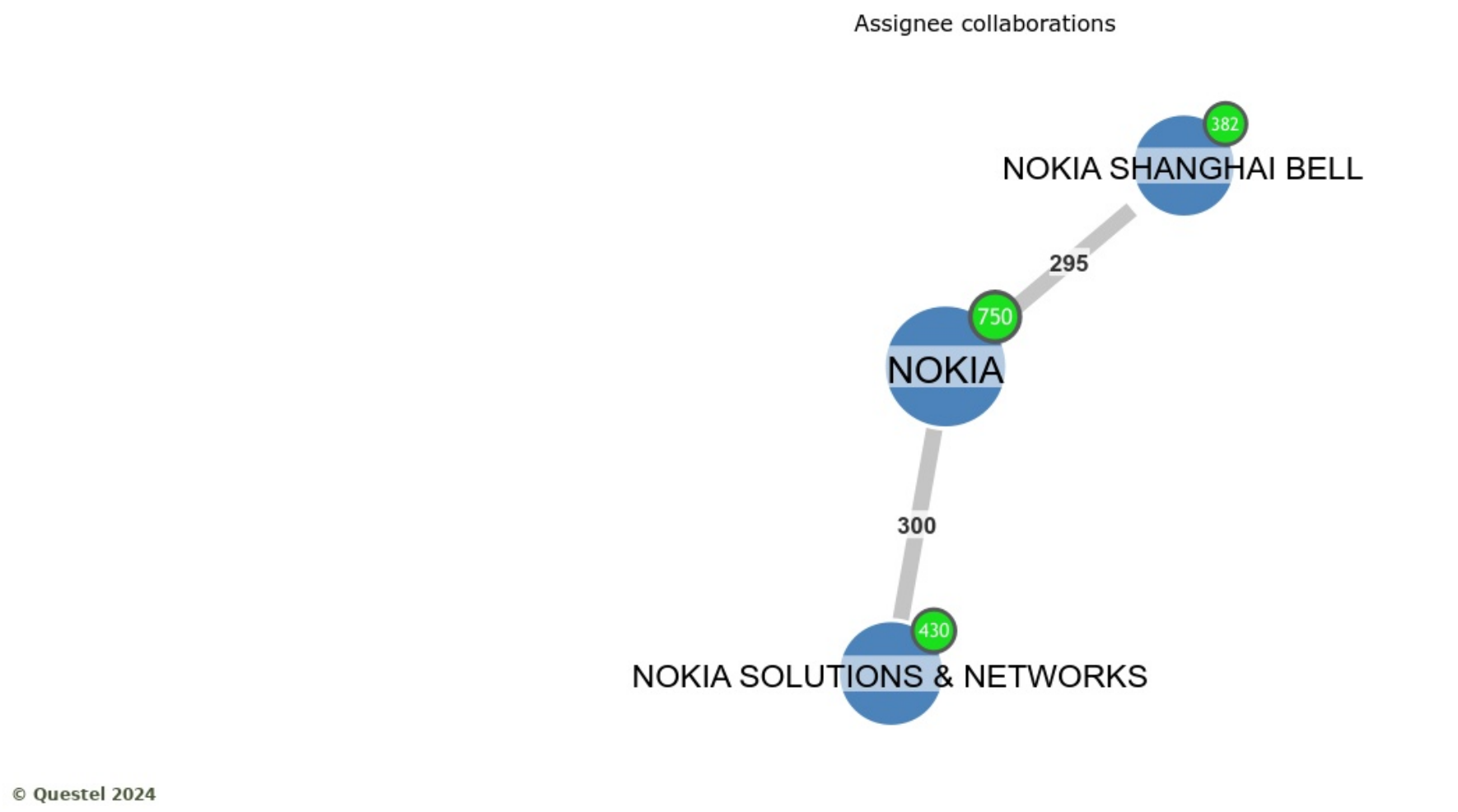

Graph 13 represents the patent collaboration network among three entities within the Nokia group: Nokia, Nokia Solutions & Networks, and Nokia Shanghai Bell. This is an intra-firm collaboration. First, Nokia serves as the central hub in this network, possessing a substantial portfolio of 750 patents. It is not just a focal point of collaboration, but also the driving force behind the group's innovation and intellectual property. Its strategic importance is evident in its substantial patent portfolio, which is the largest among the three entities.

Secondly, Nokia Solutions & Networks, a specialist in telecommunications infrastructure, holds 430 independent patents. Its connection with Nokia is not just significant, but profound, with 300 collaborative patents indicating a deeply integrated research and development effort. This collaboration signifies a concerted push to advance key technologies in the telecommunications sector. Thirdly, operating in China, Nokia Shanghai Bell independently possesses 382 patents. Its connection to Nokia is robust, with 295 collaborative patents, underscoring the entity's significant contribution to the group's innovation activities. The close numerical alignment between collaborative and independent patents suggests that a substantial portion of Nokia Shanghai Bell's innovation has conducted in partnership with the parent company. The Graph or diagram highlights Nokia's strategic focus on fostering internal collaboration across its subsidiaries. The extensive patent collaborations between Nokia and its subsidiaries reveal a continuous exchange of knowledge and technology, with each entity contributing to a cohesive and robust global patent portfolio. As the parent company, Nokia not only holds the majority of patents but also facilitates synergies across its global divisions. These collaborative model advantages regional and technical specialisations within its subsidiaries to drive shared and applied innovations across the organisation. This data fosters a culture of innovation and ensures that the best ideas and technologies are utilised across the group, enhancing its competitive edge.

**Table 3 – First 10 patents on 5G/6G and remote surgery by Citations Impact (Orbit Intelligence, 2024)**

| Title | Family Normalised Assignee Name | Publication Numbers | Earliest Application Year | Average Citation Impact | Patent strength | Geographic Coverage |
|---|---|---|---|---|---|---|
| Method of compressing tissue within a stapling device and simultaneously displaying the location of the tissue within the jaws | CILAG INTERNATIONAL | EP3545862 | 2018 | 12.86 | 7.7 | 5 |
| Network architecture, methods, and devices for a wireless communications network | ERICSSON | EP3681197 | 2016 | 11.99 | 9.5 | 10 |
| Cooperative utilization of data derived from secondary sources by intelligent surgical hubs | ETHICON | EP3506280 | 2018 | 11.81 | 8.22 | 7 |
| Communication hub and storage device for storing parameters and status of a surgical device to be shared with cloud based analytics systems | ETHICON | EP3505042 | 2018 | 11.56 | 6.99 | 5 |
| Network architecture, methods, and devices for a wireless communications network | ERICSSON | EP3499785 | 2016 | 11.55 | 9.29 | 9 |
| Surgical instrument with a tissue marking assembly | ETHICON | EP3505130 | 2018 | 11.44 | 7.72 | 7 |
| Surgical modular energy system with footer module | ETHICON | EP4360575 | 2019 | 11.31 | 6.49 | 5 |
| Cloud-based medical analytics for security and authentication trends and reactive measures | ETHICON | EP4064292 | 2018 | 11.31 | 7.82 | 7 |
| Systems, methods, kits, and apparatuses for robotic process automation in value chain networks | STRONG FORCE VCN PORTFOLIO 2019 | US20240144181 | 2022 | 11.26 | 5.64 | 1 |
| Multi-entity resource, security, and service management in edge computing deployments | INTEL | US20210144517 | 2020 | 11.24 | 5.63 | 1 |

Graphs 12 and 13 reveal a significant trend in the patent TIS, with Chinese companies, particularly Huawei and Beijing Xiaomi Mobile Software, leading in patent quantity. Notable companies include Guangdong Oppo Mobile Telecommunications and Nokia Shanghai Bell. Conversely, the table above draws attention to the significant impact of patent citations from companies not featured in the graphs. For instance, Cilag International, a Swiss pharmaceutical company associated with Johnson & Johnson, is now venturing into Advanced Medicine, particularly remote surgery using 5G/6G. This underscores the importance of considering a more comprehensive view of the patent TIS. Similarly, even though they do not appear in the patent quantity graph, companies such as Ericsson, Ethicon, and Strong Force VCN Portfolio 2019 have a more significant impact on the market with their patent citations than Chinese companies. This shows that their patents may be more important in the market in general than Chinese ones, and this is a fact of crucial importance. When it comes to quality, the patents of the companies highlighted in Table 3, such as Cilag International, Ericsson, Ethicon, and Strong Force VCN Portfolio 2019, stand out. Their patents are more relevant, significantly altering the analysis of the TIS in question.

**Graph 14 - Patents on 5G/6G and Remote Surgery at International TIS of Collaboration Network - Zooming in HHS Node**

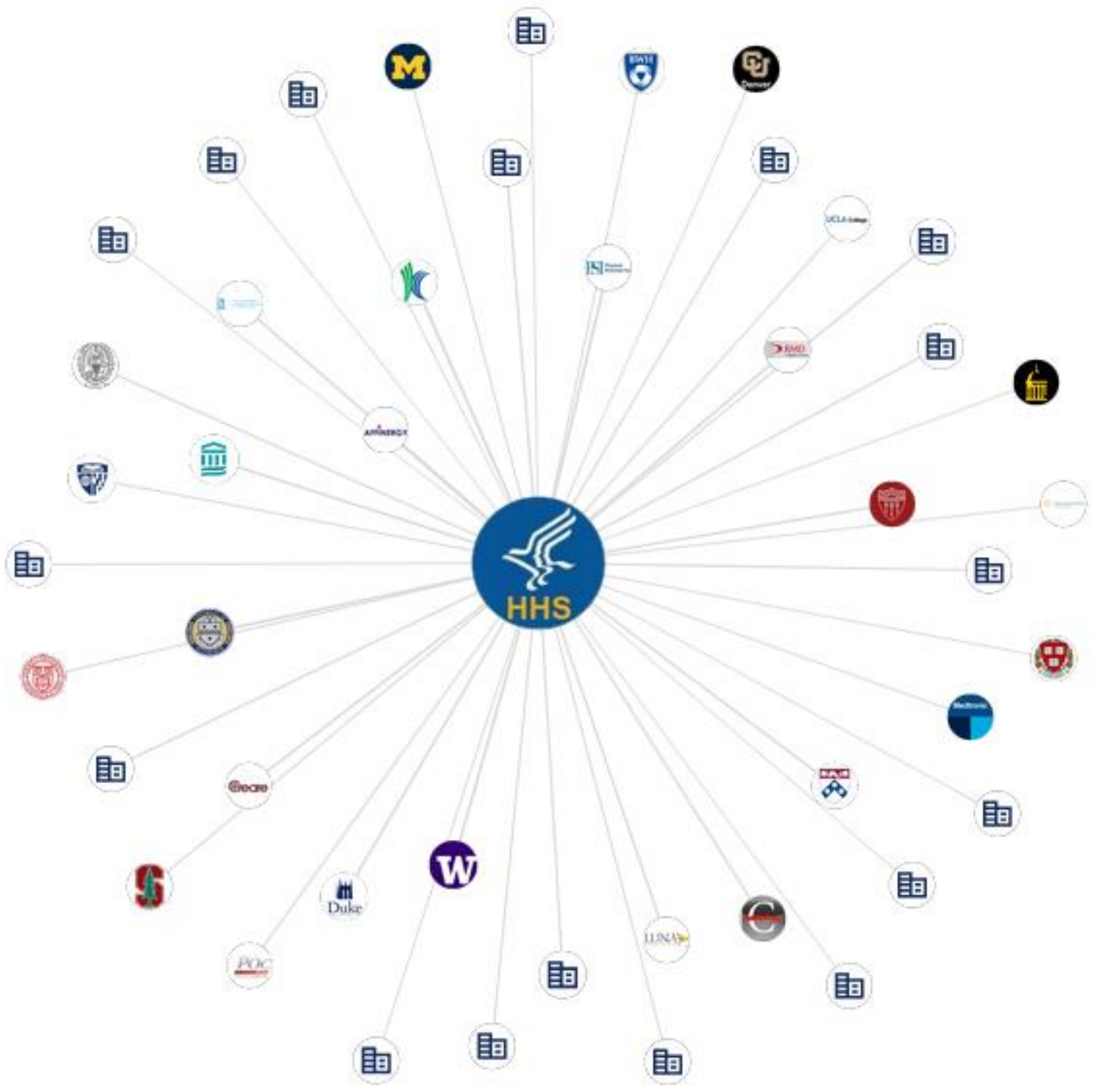

**Table 4 - Patents by Organisations on 5G/6G and Remote Surgery at International TIS (Orbit Insight, 2024)**

| Name | Rank | Acronym | Type | Country |
|---|---|---|---|---|
| U.S. Department of Health & Human Services | 1 | HHS | National/Governmental Organization | US |
| U.S. Department of Defense | 2 | | National/Governmental Organization | US |
| National Science Foundation | 3 | NSF | Government | US |
| Alcon AG | 4 | | Company | CH |
| Assiut University | 5 | | University | EG |
| United States Coast Guard | 6 | | Government | US |
| Massachusetts General Hospital | 7 | MGH | Health Care Institution | US |
| Johnson & Johnson | 8 | | Company | US |

Table 7 has been created through Orbit Insight, which generated a CSV archive to develop this metric and provide an understanding of Graph 14. Therefore, Graph 14 and Table 7 illustrate collaborative networks associated with the U.S. Department of Health and Human Services (HHS). The central node, HHS, is linked to various entities, including universities, research institutions, and companies, indicating partnerships or co-ownership in patents and research publications. Some examples of these institutions are the National Science Foundation, Alcon AG, Assiut University, United States Coast Guard, Massachusetts General Hospital, and Johnson & Johnson. Each icon or logo surrounding the central HHS node represents different collaborating institutions. Lines connecting these institutions to HHS signify active collaborations in 5G, 6G, and remote surgery—all that range from joint research projects to shared patents aimed at advancing public health innovation. The map vividly displays the extensive and diverse network of HHS, underlining its role in fostering innovation through partnerships with leading academic and research organisations. This interconnectedness, spanning various fields, is crucial for advancing medical research and developing new health technologies.

## 4.5. Sample IV - Patents at National TIS – Germany (Functions 2 and 3)

### Graph 15 - Patents on 5G/6G and Remote Surgery at National TIS by Assignees

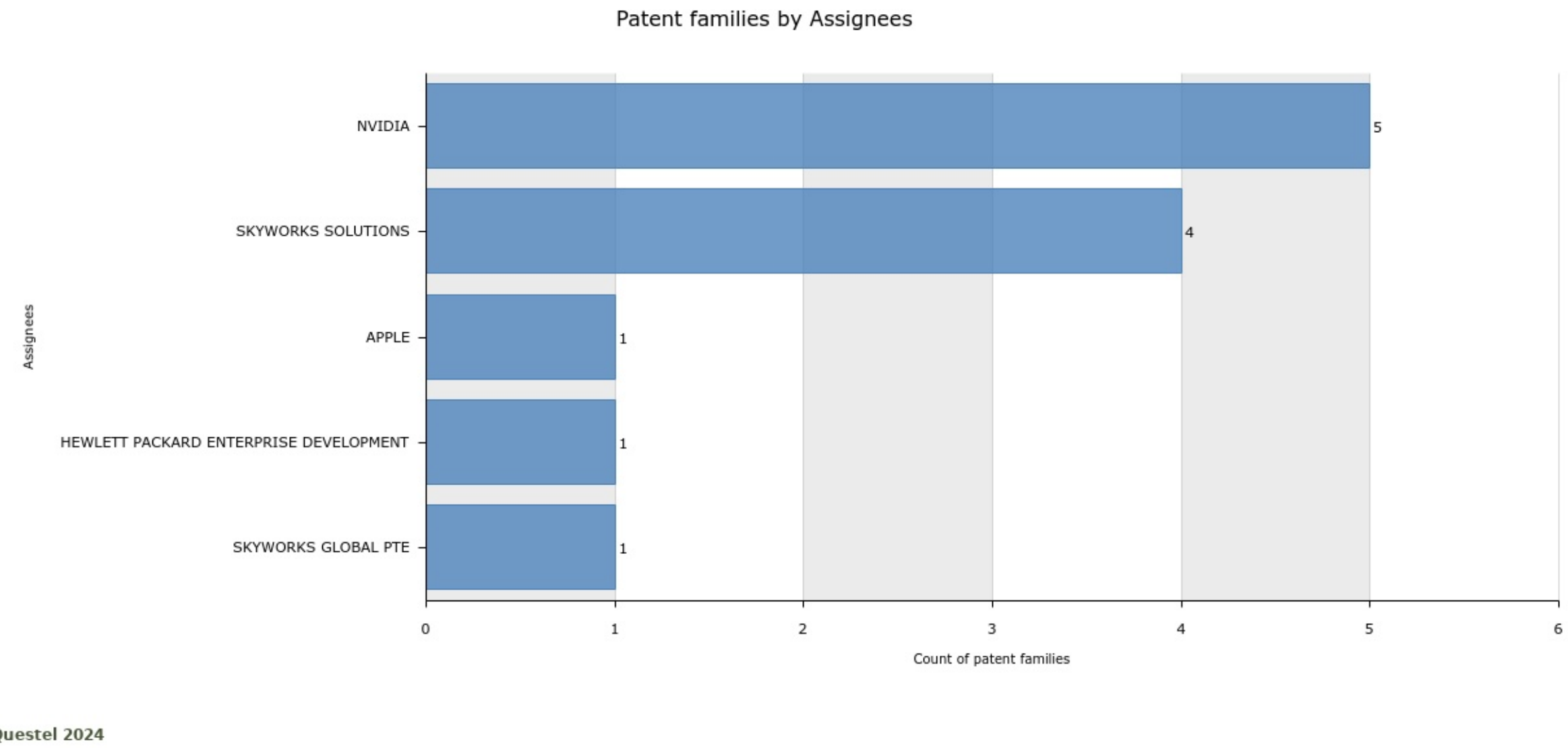

Graphic 18 illustrates the distribution of patent families ***in Germany*** among different assignees, highlighting the diversity in patent holdings. A patent family is a collection of documents covering a single invention. These companies have been prominent in Germany between 2014 and 2024. NVIDIA holds the highest number of patent families, with a total 5. Skyworks Solutions follows with four patent families. Apple, Hewlett Packard Enterprise Development, and Skyworks Global PTE hold one patent family each. This graphic highlights the leading organisations regarding patent holdings in this context, demonstrating our analysis's breadth. NVIDIA is the most dominant, suggesting a strong emphasis on innovation and intellectual property within its field. It demonstrates the importance of foreign capital in Germany, with enterprises from other countries and regions of the world dominant in the patent map. In Germany, NVIDIA is a strategic player in Generative AI, reconstitution of medical images, speaking AI, and digital biology. Skyworks Solutions, on the other hand, has a portfolio that could potentially revolutionize the medical and technology industries. Their portfolio includes semiconductors, such as low-noise and power amplifiers, connectivity switches or radio co-processors, which can be used in the automotive, streaming or broadcast industry.

Apple has the XR Vision Pro, which can manage the surgical planning interface or medical education, with a crucial definition in the health system. Hewlett Packard Enterprise Development (HP) is a North American company of printers and 3D printers, successful for remote surgery with 5G or 6G.

**Graph 16 - Patents on 5G/6G and Remote Surgery at National TIS of Collaboration Networks in Germany - Zooming in Startseite Fraunhofer-Gesellschaft Node**

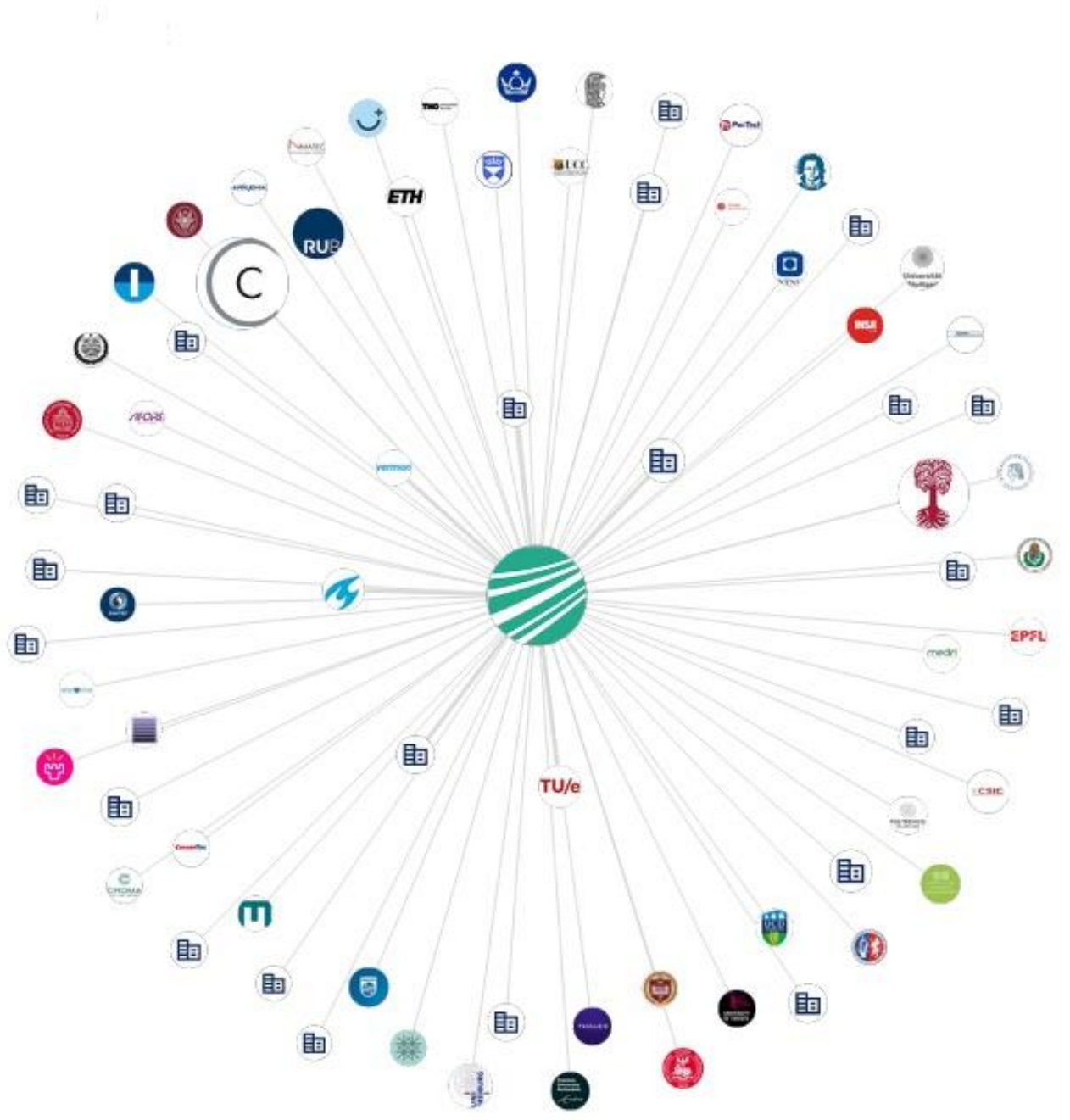

**Table 5 - Patents by Organisations on 5G/6G and Remote Surgery at National TIS (Orbit Insight, 2024)**

| Name | Rank | Type | Country | Relation | Documents |
|---|---|---|---|---|---|
| Technical University of Munich | 1 | University | DE | 99 | 119 |
| Fraunhofer-Gesellschaft | 2 | National/Governmental Organization | DE | 108 | 114 |
| Charité - Medical University of Berlin | 3 | University | DE | 90 | 100 |
| University of Tübingen | 4 | University | DE | 62 | 64 |
| University of Ulm | 5 | University | DE | 50 | 59 |
| DKFZ German Cancer Research Center | 6 | Company | DE | 48 | 55 |
| University Hospital Heidelberg | 7 | Health Care Institution | DE | 48 | 50 |
| University of Leipzig | 8 | University | DE | 46 | 49 |

Graphic 18 and Table 7 illustrate the collaborative networks between different organisations and affiliations (universities, companies and research institutes) in technological innovation system. The central node represents a key organisation (probably a prominent research institute or a major company) connected to several other entities that have contributed relevant patents or publications. The connections indicate collaborations or co-participations in patents, demonstrating the interdependence and cooperation within the 5G/6G TIS interconnected with the Remote Surgery TIS. Each icon or logo around the central node represents a specific contributing entity. Many icons suggest a robust and diverse network of partners, essential for technological advancement and innovation. Simply put, this diagram visualises how organisations in Germany have been interconnected through patents, highlighting the importance of collaboration for creating and developing new technologies in an interactive way between demand-pull and science push.

# 5. Analysis

## 5.1. Germany as a Strategic Actor in 5G/6G TIS

Understanding publications and patents, German cities like Munich, Berlin, Düsseldorf, Tübingen, Heidelberg, and Leipzig are research and development hubs. The German government may include targeted Incentive Policies for patents and publication creation and R&D activities focused on 5G and 6G in the medical area. As highlighted by Martinelli et al. (2024), it is crucial to design public policies that advance 5G and 6G, given their complementarity and interplay. Moreover, the importance of international cooperation cannot be overstated, and it should be actively promoted through targeted incentive policies. This cooperation is vital for scientific advancement and technological dissemination, particularly regarding patents and technology transfer. We propose policy measures designed to enhance interactions between smaller suppliers and larger corporations, recognising the vital role of these smaller actors as intermediaries facilitating access to international markets. If implemented by the German government, these policies could drive significant advancements, particularly in the medical system, fostering collaborative TIS where stakeholders could contribute to technological and medical innovation. The Startseite Fraunhofer-Gesellschaft in Munich focused on patents, publications, and co-participation in collaborative projects, underscores its central role and significance in the city's research panorama. Consequently, Germany, while not a dominant player, has been strategically positioned in the industrial application of 5G/6G networks for remote surgery, because of its diverse conglomerate of enterprises from around the world, including research institutes and public organisations. This strengthens Germany's position as a leader in medical innovation and benefits patients worldwide. The study of 5G and 6G networks applied to remote surgery in Germany is justified by its leadership in medical science and advanced technological infrastructure and the significant benefits that these technologies can bring to public health and the economy. Our study focuses on this context, contributing to scientific and technological advancements while influencing global health policies and innovation strategies.

# 6. Discussion & Conclusions

The pioneering advancements of 5G and 6G networks are prominent in the realms of commercialization and market development, reflecting a mature stage within the TIS. The central question of our article interplayed with the structure of our argument, highlighting Functions 2 and 3 are not just crucial, but pivotal in developing key aspects t of the Technological Innovation System of 5G. While not our primary focus, it was possible to trace especially Function 1, entrepreneurial activity, and Function 5, market formation.

Publications and patents, such as the number of published research articles or registered patents, reveal notable contributions to the technology transfer perspective. TT in Germany has been linked to a pragmatic perspective of the German National TIS regarding its ability to collaborate little internationally, often choosing to collaborate internally with institutions that are physically present on national territory. Beyond these metrics, network formation within these domains is also important. These indicators can be used to quantify how a country can produce regional innovation hubs. For instance, a high number of patents in a specific region could indicate a strong technology transfer environment that will direct and intensify the speed of innovation in a particular context, Germany or any other country. It is possible to understand market formation in many aspects and levels, whether international, national, or regional. In this case, it is unnecessary to understand regional perspectives, which would require other economics analysis. We can work with macro spheres to understand microspheres without entering microeconomic approaches. After all, understanding market formation makes it possible to comprehend how technological change is happening, at least in industrial terms. This fact only happens at advanced maturity stages of TIS. 5G TIS is not only advanced but preceding 6G TIS. Of course, it happens in some contexts, such as Europe and the United States. On the other hand, global technological change is much more integrative and dynamic in R&D terms. These changes might accelerate industrial activities. Future studies on TIS, 5G/6G and Remote Surgery could explore other TIS functions more extensively, ensuring they has been correlated with international standardisation. Furthermore, research could examine the interplay between international and national levels through empirical case studies incorporating software analysis and interviews. These future studies could overcome many limitations of the present article, such as analysing the Healthcare TIS context through patient monitoring by new technologies (5G and 6G) with an in-depth discussion about standardisation.

## Appendix (Eq. 1-A, Eq. 2-B)

### *Eq. 1-A*

| **Publications – Eq. (A1)** |
|---|
| TS=(((("5g technolog*") OR ("5g mobile*") OR ("5g wireless") OR ("5g network*") OR ("5g communication")) OR (("fifth generation technolog*") OR ("fifth generation mobile*") OR ("fifth generation wireless") OR ("fifth generation network*") OR ("fifth generation communication")) OR (("5th generation technolog*") OR ("5th generation |

mobile*") OR ("5th generation wireless") OR ("5th generation network*") OR ("5th generation communication"))) OR ((("6g technolog*") OR ("6g mobile*") OR ("6g wireless") OR ("6g network*") OR ("6g communication")) OR (("sixth generation technolog*") OR ("sixth generation mobile*") OR ("sixth generation wireless") OR ("sixth generation network*") OR ("sixth generation communication")) OR (("6th generation technolog*") OR ("6th generation mobile*") OR ("6th generation wireless") OR ("6th generation network*") OR ("6th generation communication"))) OR ((("future wireless technolog*") OR ("future wireless mobile*") OR ("future wireless network*") OR ("future wireless communication"))) OR ((("next generation technolog*") OR ("next generation mobile*") OR ("next generation wireless") OR ("next generation network*") OR ("next generation communication")))) AND TS=(“surger*” OR “surgical” or “telesurger*”)

***Eq. 2-B***

| **Patents Formula** | | | |
|---|---|---|---|
| **Selected Terms for 5G/6G** | **1** | ((5_G OR FIFTH_GENERATION OR 5TH_GENERATION OR 5_GENERATION) OR (6_G OR SIXTH_GENERATION OR 6TH_GENERATION OR 6_GENERATION)) 2W (MOBILE OR COMMUNICATION OR WIRELESS OR NETWORK? OR INTERNET OR TECHNOLOG? OR SYSTEM?)/BI/SA | **53,533** |
| **Selected Terms for 5G/6G** | **2** | ((5_G OR FIFTH_GENERATION OR 5TH_GENERATION OR 5_GENERATION) OR (6_G OR SIXTH_GENERATION OR 6TH_GENERATION OR 6_GENERATION))/BI/SA/TX/KEYW AND (H04+ OR H01Q)/IPC/CPC | **357,018** |
| **Selected Terms for 5G/6G** | **3** | ((((+HIGH+ OR +FAST+) W (SPEED)) OR ((+LOW+ OR MINIM+) W (LATENCY OR DELAY+)) OR NEXT GENERATION OR FUTURE OR CUTTING EDGE OR ADVANCED) 2W (MOBILE OR COMMUNICATION OR WIRELESS OR NETWORK? OR INTERNET))//BI/SA | **67,381** |
| **Selected Terms for 5G/6G** | **4** | ((((+HIGH+ OR +FAST+) W (SPEED)) OR ((+LOW+ OR MINIM+) W (LATENCY OR DELAY+)) OR NEXT GENERATION OR FUTURE OR CUTTING EDGE OR ADVANCED) 2W (MOBILE OR COMMUNICATION OR WIRELESS OR NETWORK? OR INTERNET))/BI/SA/TX/KEYW AND (H04+ OR H01Q)/IPC/CPC | **33,1068** |
| **Selected Terms for Robotic Surgery Classification** | **5** | (((REMOTE OR MINIMALLY INVASIVE OR NON_INVASIVE OR ROBOT+ OR TELE+) 2W (+SURGER+ OR SURGICAL)) OR TELESURG+)/BI/SA/TX/KEYW | **89,601** |
| **Robotic Surgery Classification** | **6** | A61B-034+/IPC/CPC | **42,898** |
| **Time horizon** | **7** | EPRD>=2014 | **38,630,137** |
| **Intersystemic Perspective** | **8** | (1 OR 2 OR 3 OR 4) AND (5 OR 6) | **10,798** |
| **Intersystemic Time Limited** | **9** | (1 OR 2 OR 3 OR 4) AND (5 OR 6) AND 7 | **10.679** |
| **International TIS – Result 2** | **10** | **(1 OR 2 OR 3 OR 4) AND (5 OR 6) AND 7 AND NBPC >1** | **5,704** |
| **National TIS – Result 3** | **11** | **(1 OR 2 OR 3 OR 4) AND (5 OR 6) AND 7 AND DE/PN** | **51** |
| **Regional TIS - Europe** | **12** | **(1 OR 2 OR 3 OR 4) AND (5 or 6) AND 7 AND EP/PN** | **2,744** |